\documentclass[sigconf,nonacm]{acmart}

\usepackage{longtable}
\usepackage{hyperref}
\usepackage{natbib}
\usepackage{wrapfig}
\usepackage{array}
\usepackage{comment}
\usepackage{float}
\usepackage{placeins} %
\AtBeginDocument{%
  }

\setcopyright{none}

\newcommand{\quot}[1]{\textit{``#1''}}

\begin{document}

\title[TouchingSpace: An Audio-Haptic Map for Blind and Low-Vision Readers]{\texorpdfstring{``I didn't know how to read a map, but now I can''}{“I didn't know how to read a map, but now I can”}: TouchingSpace, an Audio-Haptic Map for Blind and Low-Vision Readers}

\author{Li Liu}
\email{lliu112@ucsc.edu}
\orcid{0000-0002-5184-054X}
\affiliation{%
  \institution{University of California Santa Cruz}
  \city{Santa Cruz}
  \state{California}
  \country{USA}
}

\author{Yihe Wang}
\email{ywan1125@ucsc.edu}
\orcid{0000-0003-3469-1359}
\affiliation{%
  \institution{University of California Santa Cruz}
  \city{Santa Cruz}
  \state{California}
  \country{USA}
}

\author{Jiaming Qu}
\authornote{Work unrelated to position at Amazon.}
\email{qjiaming@amazon.com}
\orcid{0000-0003-4460-5637}
\affiliation{%
  \institution{Amazon}
  \city{Seattle}
  \state{Washington}
  \country{USA}
}

\author{Ashmita Dua}
\email{asdua@ucsc.edu}
\orcid{0009-0007-8337-7848}
\affiliation{%
  \institution{University of California Santa Cruz}
  \city{Santa Cruz}
  \state{California}
  \country{USA}
}

\author{David T. Lee}
\email{dlee105@ucsc.edu}
\orcid{0000-0002-1495-7052}
\affiliation{%
  \institution{University of California Santa Cruz}
  \city{Santa Cruz}
  \state{California}
  \country{USA}
}

\author{Leilani H. Gilpin}
\email{lgilpin@ucsc.edu}
\orcid{0000-0002-9741-2014}
\affiliation{%
  \institution{University of California Santa Cruz}
  \city{Santa Cruz}
  \state{California}
  \country{USA}
}

\renewcommand{\shortauthors}{Liu et al.}

\begin{abstract}
Accessible map systems either make a layout explorable by hand or convey information through speech; few combine both to support pre-travel spatial understanding for blind and low-vision (BLV) people. We present TouchingSpace: a system that retrieves map data for an outdoor place and renders its surroundings as bounded regions at fixed trackpad positions. During exploration, users receive audio and haptic feedback and can ask a conversational agent open-ended questions. We conducted a user study with fourteen BLV participants who explored a place using TouchingSpace and reflected on the experience. We found participants used sound and vibration to locate places and speech to identify and describe them; the bounded surface supported discovery, revision, and spatial checks by hand; they expected this awareness to support future travel. TouchingSpace demonstrates how a laptop trackpad can support self-directed spatial exploration. These findings suggest accessible AI maps should ground conversation on bounded, user-controlled spatial surfaces.
\end{abstract}



\begin{teaserfigure}
  \includegraphics[width=\textwidth]{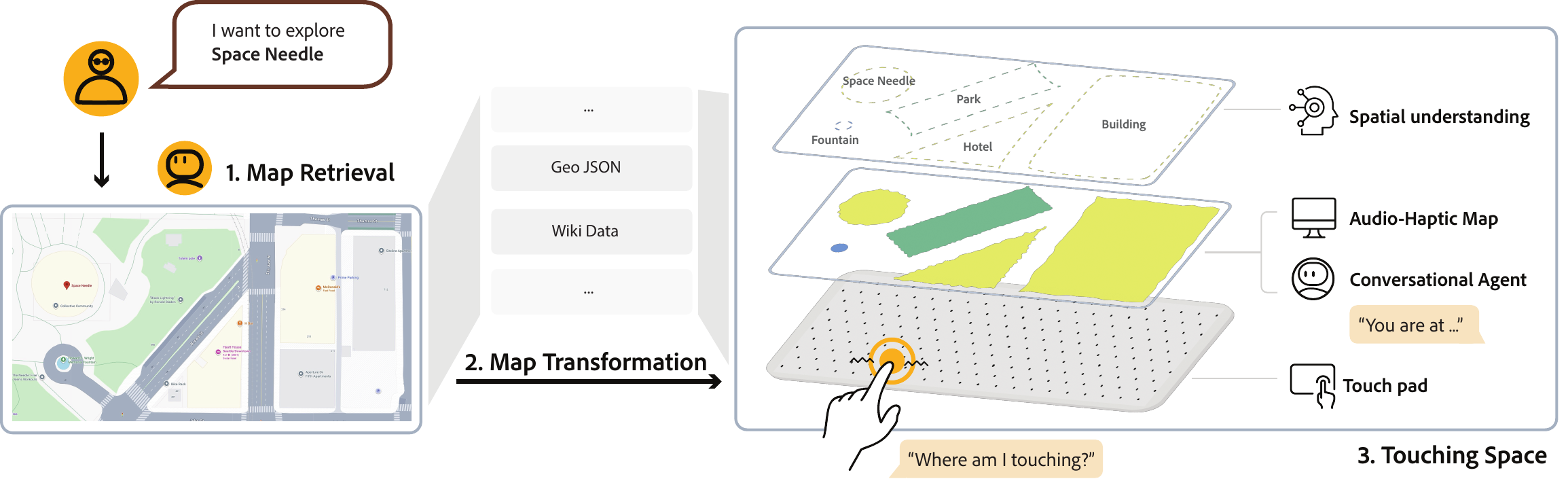}
  \Description{A three-stage pipeline. Stage 1 shows a spoken request for a place, a map tile, and a data stack labeled Geo JSON and Wiki Data. Stage 2 shows the raw geographic data transformed into colored polygon zones on a map. Stage 3 shows a reader exploring the audio-haptic map on a trackpad while a conversational agent answers questions, with the resulting spatial understanding, audio-haptic map, conversational agent, and touch pad labeled.}
  \caption{TouchingSpace pairs a touchable map with a conversational agent so a blind or low-vision reader can read a place before a trip and ask about it. (1) The system retrieves map data for a named place. (2) It renders the geographic data as an audio-haptic map. (3) The reader explores the map by finger on a trackpad while the agent answers questions about the place under the finger.}
  \label{fig:teaser}
\end{teaserfigure}

\maketitle

\section{Introduction}
Maps help travelers understand unfamiliar places before visiting them. Beyond showing a route, a map reveals which places are present, what they are, and how they are arranged relative to one another~\cite{thorndyke1982maps}. Most maps make this information readily available to sighted readers, but Blind and low-vision (BLV) travelers often acquire spatial information through other approaches: traveling along routes, listening to turn-by-turn directions, or asking questions about particular destinations. However, a route learned as a sequence of turns does not necessarily convey the layout around it~\cite{thinusblanc1997representation}. Moreover, asking about a place requires knowing that the place exists. As a result, BLV travelers may remain unaware of relevant places even along routes they frequently travel~\cite{banovic2013uncovering}. Access to an overview of an unfamiliar place therefore matters not only for reaching a known destination, but also for understanding its surroundings and discovering places one did not already \mbox{know~to~ask~about.}

Prior work has supported BLV travelers' spatial understanding through two approaches. The first makes maps touchable. Tactile maps make spatial arrangements explorable by hand~\cite{brock2015interactivity,ducasse2018tangible}; vibro-audio maps on flat touchscreens can support map learning as effectively as matched embossed maps~\cite{palani2022comparing}; and Apple Maps with VoiceOver speaks the label beneath a moving finger~\cite{applemapsvoiceover}. These interfaces provide a persistent representation of \emph{where} places are, but the places and descriptions available to the reader are generally determined in advance. The second approach uses camera-based services, voice assistants, and street-view agents to answer open-ended questions about a photograph, a named place, or a selected panorama~\cite{bigham2010vizwiz,abdolrahmani2018siri,froehlich2025streetviewai}. These systems provide richer access to \emph{what} a place is, but present information sequentially and without a persistent spatial reference. Their answers can therefore be difficult to assess: BLV users of a street-view agent rated its answers as accurate while also reporting difficulty determining whether an answer was correct~\cite{froehlich2025streetviewai}. Soundscape and its successor can announce places that a traveler did not name, but they do so during travel rather than through a pre-travel overview~\cite{microsoft2018soundscape,voicevista}.

Prior studies provide complementary but disconnected information about a place: touchable maps preserve the \emph{where}, whereas conversational systems elaborate the \emph{what}. To bring both into one self-directed reading, we developed \emph{TouchingSpace}, a trackpad-based audio-haptic map combined with a conversational agent. We developed the system with a blind co-designer whose feedback shaped its core functions. Given an outdoor place, TouchingSpace retrieves geographic data, renders nearby places as bounded regions, and maps them to fixed positions on a haptic trackpad. To manage information density, the system reveals the map in four cumulative stages, progressing from a single central anchor to the full map. During exploration, users receive both audio and haptic feedback: a proximity sound grows stronger as the finger approaches a place; crossing its boundary causes the trackpad to vibrate and the system to speak its name. While resting on a place, the user can ask open-ended questions about what it is, or about spatial relations such as its direction or distance to another place. The agent's answers are grounded in the map geometry, the place under the finger, and the recent exploration and conversation history.\looseness=-1

To understand how TouchingSpace helps BLV people build spatial knowledge of an unfamiliar place before traveling to it, we conducted an in-person study with fourteen participants recruited through a local accessibility center. Each participant first learned the system on a practice map and then freely explored a personally relevant outdoor place while thinking aloud. Participants then completed an exit interview reflecting on their experience. We analyzed their finger trajectories, key presses, and questions alongside their think-aloud comments and interview responses. Our study addressed three research questions (RQs):

\begin{itemize}
 \item \textbf{RQ1.} What information did participants obtain from the audio, haptic, and conversational modalities, and how did they combine it?
 \item \textbf{RQ2.} How did participants discover, revise, and check their understanding of a place while exploring the map?
 \item \textbf{RQ3.} How did participants expect to use what they learned when preparing for future travel?
\end{itemize}

We answer these RQs through three main findings. First, participants took a place's \emph{where} from the proximity sound and vibration and its \emph{what} from spoken names and the agent, while the agent also provided map-relative directions. They read these complementary modalities against one another when information was missing or uncertain (RQ1). Second, participants used the bounded, absolutely mapped surface to organize places around spatial anchors. On this surface, they discovered places they had not asked about, revised their prior understanding, and checked spoken spatial relations by hand (RQ2). Third, participants expected to use this surrounding awareness before travel to plan routes and fallbacks, incorporate newly discovered destinations into trips, and complement the navigation tools they already use on the street (RQ3).

In summary, this paper makes two contributions. First, we present TouchingSpace, an accessible map system that combines an audio-haptic map with a conversational agent whose answers are grounded in the map geometry and the reader's current interaction state. Second, we provide empirical evidence of how BLV people use conversation grounded in an audio-haptic map to build spatial understanding and discover places they had not known to ask about. Our findings suggest a broader design implication for AI-enabled accessible maps: \textbf{conversational depth should be combined with a bounded spatial representation under the reader's control}. Such a representation lets BLV people plan the order and density of exploration ahead, and encounter unanticipated places and check spatial claims against the same surface.\looseness=-1

\section{Related Work}

\subsection{Spatial knowledge without sight}

Spatial knowledge is commonly described at three levels: landmarks, the routes between them, and the survey level that holds them in one layout~\cite{siegel1975development,schinazi2016navigation}. \emph{Landmarks} are the places a traveler recognizes and uses as reference points, one place at a time. A few places, not always public landmarks, are held to anchor the rest of what a sighted traveler knows of an area~\cite{couclelis1987anchor}. \emph{Route knowledge} gets a traveler from one place to the next, held as a sequence of instructions~\cite{thinusblanc1997representation,kitchin1997understanding}. \emph{Survey knowledge} is the layout itself, the distance and direction among the landmarks a traveler holds.

Blind and low-vision travelers build that layout out of what they gather on their own routes and from asking others~\cite{williams2013pray}. However, both sources have limits. Their own routes leave places out: participants reported places they cared about going unknown for as long as two years~\cite{banovic2013uncovering}. Asking needs a question, and a traveler cannot ask about a place they do not know exists~\cite{belkin1995cases}. Blind web users face the same gap: they often cannot tell whether information is inaccessible or simply absent~\cite{bigham2017nkwydk}. Places learned on separate trips may also stay uncoordinated, which is reported of sighted travelers as well~\cite{montello1998framework}. We study what a blind reader does with a whole layout, on a map they can touch and question before a trip.

\subsection{Audio and speech assistance for conveying spatial information}

Sound carries a place to a blind reader in several forms. A non-speech sound stands for something a name would have to say: which way a finger must move to regain a shape~\cite{su2010timbremap}, or how near a walker is to a waypoint~\cite{wilson2007swan}. A spoken name gives a place its identity: VoiceVista, an independently maintained successor to Microsoft's discontinued Soundscape~\cite{microsoft2018soundscape}, announces nearby landmarks, streets and intersections as a walker passes~\cite{voicevista}, and Audiom speaks what an avatar meets as a reader moves it through a map with the arrow keys~\cite{biggs2022evaluation}. And an answer comes back when a reader asks: about where they are and what is near them~\cite{abdolrahmani2018siri}, about a region a reader has stepped to with the keyboard~\cite{geovisa11y2026}, or about a street view read before a trip~\cite{froehlich2025streetviewai}.

Non-speech sound, spoken names, and answers each carry different spatial information, and the listener has to combine them alone. Turning a two-dimensional display into a linear list of targets makes it accessible but removes the arrangement it had~\cite{kane2011access}. Blind and low-vision travelers who navigate by smartphone carry three or four navigation apps and switch among them for different levels of feedback~\cite{williams2013pray}. Systems usually bring one or two of these forms together, such as a beacon with spoken callouts~\cite{voicevista}. We ask what a blind reader takes from each form when all three are on one map of a place they chose.

\subsection{Touchable maps and the layout under the hand}

A touchable map gives a reader a persistent layout under the hand. It may be a raised-line or 3D printed sheet~\cite{holloway2018accessible,nagassa2023building}, a plain touchscreen that vibrates and speaks~\cite{giudice2012learning,poppinga2011touchover}, a plan explored with a gamepad~\cite{kaplan2024audiohaptic}, or a room explored through a force-feedback joystick~\cite{lahav2008haptic}. Readers reconstruct a map from a vibrating touchscreen about as well as from a matched tactile sheet~\cite{palani2022comparing}, and walk an indoor layout about as accurately as after an embossed overlay~\cite{giudice2020cognitive}. Blind and low-vision readers prefer to receive such a map at home and learn it in advance of a trip~\cite{rowell2003feelingyour,rowell2005feeling}, and adults walked a city route about as well after studying only a tactile map as after learning it on foot~\cite{espinosa1998comparing}.

A surface shows where places are, and a label says which place it is. Braille carried those labels, and since a 2006 platform many systems have spoken them instead~\cite{cole2021tactile}. Systems attach them differently: a grid of zones~\cite{simonnet2019comparing}, audio labels on a raised-line map~\cite{brock2015interactivity,ducasse2018tangible}, a plate laid over the screen~\cite{kane2013touchplates}. Apple Maps speaks a place's label under a dragged finger~\cite{applemapsvoiceover}, and recent work has readers speak as well as touch, on maps and other touchable graphics~\cite{zhao2024tada,reinders2020heymodel,cavazosquero2019jido,jiang2026electrotactile,reinders2024rtd}. The closest to our own is MapIO, a language model behind a physical tactile map with an overhead camera tracking the pointing hand~\cite{mapio2025}. A surface made in advance fixes which places are on it. Through a digital haptic map, we explore how a reader makes sense of what they touch, and how they combine touch with speech and with their own control of what the map shows.

\subsection{Trusting and verifying a spoken account}

A spoken answer about a place cannot be visually checked, so a reader takes it on trust. The same difficulty shows up across captions, recognizers, generative tools and street-view agents: a wrong answer is taken with the confidence of a right one~\cite{macleod2017captions,adnin2024king,froehlich2025streetviewai,scenescout2026}. Where blind and low-vision participants took a recognizer to be right, they were almost as certain of the wrong outputs as of the right ones~\cite{hong2024errors}, and when a low-stakes answer is wrong, blind users of image description tools often blame their own photograph rather than the model~\cite{sakib2026xai}. How far that trust should go depends in part on how well it matches what a system can actually do~\cite{lee2004trust,hoff2015trust}, and the designs that most reduce overreliance are the ones crowdworkers on a visual task rated least favorably~\cite{bucinca2021trust}.

A reader who wants to check a spoken answer has to look somewhere else. They try the machine on something they already know, turn to another sense, ask a sighted person, or run a second device, though a second app's reading does not always settle the question~\cite{alharbi2024misfitting}. Asked how they would react to a navigation aid's error, some visually impaired respondents said they would put the device away without deciding whether it was at fault~\cite{abdolrahmani2017embracing}. Each of these brings in another person, another device, or another trip, and by then the answer is gone. We explore what a blind reader does when the claim and the check are on the same surface.

\section{The TouchingSpace System}
\label{sec:system}

\subsection{System Overview}
\begin{figure*}[t]
\centering
\includegraphics[width=\textwidth]{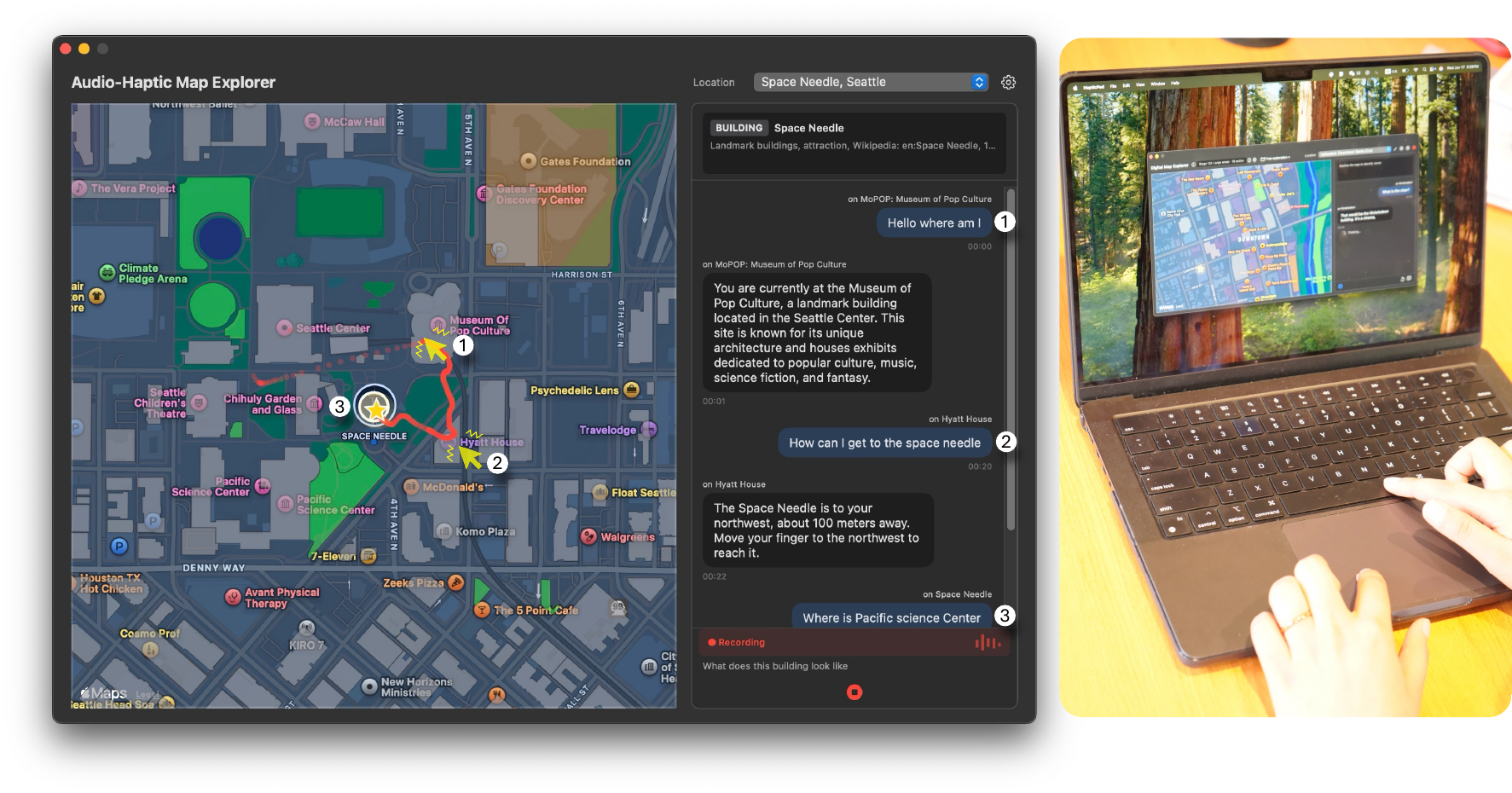}
\caption{Space Needle on TouchingSpace, as the screen shows it to sighted people; a reader takes the map by finger on the trackpad and hears the agent. Left, the map with the finger's path in red and the current position circled; the numbered flashes are the places under the finger when each question was asked. Middle, the conversation: (1) on the Museum of Pop Culture, ``Hello, where am I,'' answered with the place and what it is; (2) on the Hyatt House, ``How can I get to the Space Needle,'' answered as a direction and distance from the finger; (3) on the Space Needle, ``Where is Pacific Science Center.'' The recording bar at the bottom shows a question being spoken while the space bar is held. Right, the system in use, one hand on the pad and the other by the keys.}
\Description{Two parts side by side. Left, a screenshot of the TouchingSpace window: a map of the Space Needle area with a red finger path, a circled current position, and three numbered markers, beside a chat panel holding three questions and the agent's spoken answers, with a red recording bar at the bottom. Right, a photograph of the system in use, a laptop with one hand on the trackpad and the other by the keys.}
\label{fig:system}
\end{figure*}

TouchingSpace is a macOS application that helps BLV people learn the spatial layout of an unfamiliar outdoor place before traveling to it. It runs on a unmodified laptop with the built-in haptic trackpad, three keys on the keyboard, and the speakers. It converts geographic data into a set of touchable regions and renders them as an audio-haptic map. Users explore the map by moving a finger across a haptic trackpad, and they can verbally ask a conversational agent about the place. As the finger moves, TouchingSpace provides both audio and haptic feedback (Figure~\ref{fig:system}).

\subsection{Design Process}
\label{sec:design-process}
We developed TouchingSpace with a blind co-designer (P01), who joined as a member of the design team~\cite{ladner2015design}. P01 is in their seventies, has been blind since the age of twenty, and travels by \quot{mental maps}. With age and hearing loss, their spatial awareness has been declining, and their mental maps had gradually mismatched the ground, so they looked for tools to calibrate the mental map with the real world.
We first introduced Gemini Live\footnote{\url{https://gemini.google/overview/gemini-live/}}, which answers questions about whatever is in front of the camera at that moment; walking with it across a campus they knew well, they \quot{got totally disoriented}. When we traced the campus layout on their palm, they told us it \quot{completely overturned} the mental map they had held of the place for years: the assistant could offer only \quot{the line between point and point, and can hardly give the area} --- for them a fatal \quot{blind spot}.

P01 therefore asked for three things: maps they could touch, because touch supplies what
\quot{language can hardly make up for}; maps they could ask while reading it, drawing on online sources to say what a building is for; and maps of places they chose themselves, some unfamiliar and some already known, so they could learn new places and check old ones against their mental map. We built the system with P01 in an iterative process over fourteen months --- by email, by phone, and in sessions at their home and on campus (Appendix~\ref{app:codesign}). It progressed from concept to working prototype, and from simple synthetic shapes to real-world map. In each round, we revised the interaction experience and improved the spatial accuracy. Five design goals (DG) came out of these exchanges.
\begin{itemize}
  \item \textbf{DG1: Any place, on a mainstream device.} A printed tactile map holds one place and is fixed once made. Our co-designer wanted maps of several places, some familiar and some not, and held that a mainstream product with \quot{a little adjustment} serves the broader blind community. TouchingSpace therefore builds the map for any named place from public map data, on an unmodified \mbox{laptop~trackpad.}
  \item \textbf{DG2: Coarse first, finer as the reader asks.} Too much at once, in our co-designer's words, \quot{is no different from none.} TouchingSpace therefore opens with the few places that give an area its shape and adds smaller ones in later stages. The agent answers each question in a sentence or two and gives more only when asked.
  \item \textbf{DG3: One fixed frame under the hand.} In the first pilot the finger acted as a relative pointer, and our co-designer said, \quot{The whole point is I have no idea where my finger is in relationship to the whole map.} TouchingSpace therefore holds a whole place on the surface at once, each point of the surface a fixed position on the map.
  \item \textbf{DG4: Answer about the place under the finger.} Our co-designer wanted to ask about a place while reading it: a name or a building code alone gave them no picture of what it was, and they had asked the camera assistant for more context. TouchingSpace therefore accepts a spoken question at any time and answers about the place the finger is on. The agent responds using the underlying map data and web search functions.
  \item \textbf{DG5: Do not assume one kind of blind reader.} Our co-designer warned that causes of blindness and the senses left to a reader differ, and that they themselves had also lost hearing. TouchingSpace therefore marks a place in more than one modality: a proximity sound, \mbox{vibration,~and~speech.}
\end{itemize}

\subsection{System Functionality}
\label{sec:system-frontend}

\textbf{Map preparation.} Given the name of an outdoor place, TouchingSpace retrieves geographic data through the Overpass API\footnote{\url{https://wiki.openstreetmap.org/wiki/Overpass_API}} to cover a 400-meter radius of the target location (roughly a five-minute walking distance). We process the raw data with OSMnx~\cite{boeing2017osmnx}, converting polygon geometries into haptic zones on a fixed-scale canvas. The polygons are overlaid on an Apple Maps tile layer based on longitude and latitude coordinates. Both map layers are used as visual prompts for the conversational agent.

\textbf{Progressive disclosure.} Following DG2, TouchingSpace reveals each map in four cumulative stages: an overview with a single central anchor, the largest areas on the map, the prominent landmarks, and the full map. It assigns each place to a stage by footprint, centrality, and whether it is named, with per-stage caps that scale with the map's density so that dense and sparse maps stay legible. %
When the stage changes, TouchingSpace announces the new stage and gives an overview of the places now available.

\textbf{Audio-haptic exploration.} Users explore with one finger on the trackpad: the whole surface holds the whole place, and the edges of the trackpad are the edges of the map (DG3). Because of the absolute mapping, lifting the finger and setting it down on the same spot of the trackpad returns to the same location on the map. When the finger is outside a revealed place, a proximity hum grows louder as the finger approaches it. Crossing a place boundary starts a continuous vibration and announces the place name (DG5). Leaving the place stops the vibration.  Because the trackpad's vibration strength is fixed in hardware, a quiet tone accompanies every vibration for readers who cannot reliably feel it.
Users perform the remaining actions on a keyboard with three keys, with a tactile marker on the middle key so the set can be found by touch. Users press the right key to reveal the next disclosure stage and the left key to return to the previous stage (DG2). To ask the agent a question, users hold the middle key while speaking and release it to submit; a quick tap replays the agent's last response.

\textbf{Conversational agent.}
\label{sec:agent-competence}
Users can ask the agent spoken questions at any moment. TouchingSpace assembles a multimodal prompt carrying the transcribed question; the current location on the map; the revealed places, ordered nearest to farthest with the direction and distance of each from the finger; recently visited places, recent conversation turns; and an image of the map annotated with a compass and the finger's location. A system prompt instructs the agent to give concise spoken responses and to describe places in non-visual language from the finger's position (DG4). Directions are cardinal by default (e.g., \quot{the theater is about 400 feet to your northeast}) and page-relative if the user changes the setting (\quot{to your upper right}). The agent is told to say when its context is insufficient rather than guess, and it can call web search for questions it cannot answer with only the map data.

The agent is a hosted, general-purpose model reached through its public API, so TouchingSpace is model-agnostic: the design work sits in the multimodal prompt and the map context. We used Gemini 2.5 Flash Preview Native Audio Dialog, pinned to the 09-2025 build, because it accepts and returns speech, reads the map image, calls web search, and handles multilingual questions~\cite{gemini25}. To check that participants would hear answers grounded in the map rather than hallucinations, we benchmarked the deployed configuration before the study: we put spoken questions to it over three of the study maps, and scored its answers against references derived from the map geometry. It matched the reference on 83.0\% of 393 scored questions, and was at its most accurate on whether a place is on the map at all (98\%) and on questions of direction and distance from the place under the finger (95\% and above). The question set, the scoring rules and the per-type results are in the supplementary material.

\section{User Study}
\label{sec:study}

\subsection{Study Overview}
To investigate how people use TouchingSpace to learn about a place, we conducted an in-person study with 14 BLV adults (P02--P15; $M=7$, $F=7$), a sample size comparable to prior user studies on accessible maps~\cite{giudice2020cognitive,kane2013touchplates,reinders2024rtd,nagassa2023building}. Participants' ages ranged from 23 to 81 ($Mean = 50.31$, $S.D. = 18.94$). Eight participants described themselves as totally blind. Eleven participants had acquired vision loss and three had congenital vision loss (Table~\ref{tab:participants}). We recruited participants through a local accessibility center and its professional network. During the study, participants completed a background interview, learned to use the system, freely explored a personally relevant place, and reflected on their experience. We allowed participants to proceed at their own pace, as they varied in vision conditions, technical experience, and the time needed to become comfortable with the system. Sessions lasted 60--180 minutes ($Mean = 120$), and participants were compensated US\$20 per hour. The study was approved by our Institutional Review Board (IRB).\looseness=-1

\begin{table*}[hbtp!]
\centering
\small
\caption{Participant characteristics for the user study. P01 was the co-designer excluded from the evaluation sample; therefore, study participant IDs begin with P02. P09 completed the study in Spanish through an interpreter; all other participants completed the study in English. Several participants interacted with the conversational agent in Arabic, Cantonese, French, Italian, and Spanish.}
\begin{tabular}{>{\raggedright\arraybackslash}p{0.04\linewidth}
>{\raggedright\arraybackslash}p{0.04\linewidth}
>{\raggedright\arraybackslash}p{0.04\linewidth}
>{\raggedright\arraybackslash}p{0.18\linewidth}
>{\raggedright\arraybackslash}p{0.10\linewidth}
>{\raggedright\arraybackslash}p{0.42\linewidth}}
\toprule
\textbf{ID} & \textbf{Age} & \textbf{Gen.} & \textbf{Vision description} & \textbf{Onset} & \textbf{Current sources of spatial information} \\
\midrule
P02 & 60 & F & Totally blind & Acquired & Sighted guides; contacting destinations in advance \\
P03 & 42 & F & Totally blind & Acquired & Google Maps; BlindSquare \\
P04 & 30 & M & Blind (light perception) & Congenital & Counting streets; Siri; camera-based AI \\
P05 & 40 & F & Low vision & Acquired & Siri directions; route planning in advance \\
P06 & 51 & F & Totally blind & Acquired & Siri; Be My Eyes; asking other people \\
P07 & 45 & F & Low vision & Congenital & Orientation and mobility training; practice visits \\
P08 & 48 & F & Totally blind & Acquired & Calling ahead; BlindSquare; asking other people \\
P09 & N/A & M & Totally blind & Acquired & Calling a sighted person; camera-based AI \\
P10 & 79 & F & Totally blind & Acquired & Sighted guides; tactile and environmental cues \\
P11 & 27 & M & Totally blind & Congenital & AI assistants; Apple Maps; landmark sequences \\
P12 & 55 & M & Blind & Acquired & Transit or police services; GPS; counting blocks \\
P13 & 23 & M & Legally blind & Acquired & Google Maps; Be My Eyes \\
P14 & 73 & M & Totally blind & Acquired & Arranging a greeter; asking other people \\
P15 & 81 & M & Blind & Acquired & Sighted guides; ChatGPT \\
\bottomrule
\end{tabular}
\label{tab:participants}
\Description{A six-column table of the fourteen participants: identifier, age, gender, vision description, whether vision loss was congenital or acquired, and the sources of spatial information each currently uses.}
\end{table*}

\subsection{Study Design and Protocol}
\textbf{Study Design:} During recruitment, participants were invited to nominate an outdoor place that they planned to visit or wanted to learn about. Eight nominated a place; the other six selected one from a pool that included well-known tourist attractions and the area surrounding an accessibility center. We prepared a TouchingSpace map for each place before the session. We used personally relevant places because our goal was to understand participants' exploration and spatial-learning processes rather than compare task performance on a common map. Personal relevance also encouraged participants to pursue questions based on their own interests. When possible, they could also relate the system's information to their prior knowledge.

Our study setup consisted of a laptop running TouchingSpace, a standalone trackpad with a raised border, and a separate three-key keypad with tactile labels. All members of the research team are sighted. To reduce the risk that the protocol assumed a sighted view, our BLV co-designer (P01) ran pilot sessions \mbox{and~helped~revise~it.}

\textbf{Study Protocol:} First, after providing informed consent, including for photographs, video, audio and screen recording, participants completed a pre-task interview. The interview asked about participants' (1) vision condition and how they reason about direction; (2) prior experience with tactile and audio maps, touch-based devices, and AI assistants; and (3) current practices for learning about unfamiliar places. 

Second, participants completed a guided onboarding activity using a map of the Space Needle and its surrounding area. Participants were instructed to use the trackpad, interpret the audio-haptic feedback, move among the map's levels of detail, and interact with the conversational agent. Third, participants freely explored their selected map. We did not assign a target-finding task; participants decided where to explore, when to change the level of detail, and what to ask the agent. Participants were encouraged to think aloud during exploration.

Finally, participants completed an exit interview organized into two parts. The first part asked about \emph{constructing spatial understanding}. Participants described their mental representation of the map, identified the anchors and spatial relationships that organized it, and reflected on how that understanding developed. We also asked about the effects of progressive disclosure and the different feedback modalities on that understanding. The second part asked about \emph{coordinating touch and conversation}. Participants described their exploration and questioning strategies first, and then explained how they cross-checked information across touch, audio, and the conversational agent. They also discussed the system limitations they encountered, as well as their suggestions for improvement. The complete interview guide is available in the supplementary material.

\subsection{Data Collection and Analysis}
We collected two types of data during the study. \emph{Behavioral data} were automatically logged by the system and included timestamped finger positions, key presses, and cursor trajectories. \emph{Verbal data} included participants' spoken queries, think-aloud comments, and interview responses. Sessions were audio- and screen-recorded, transcribed, and de-identified. We used Whisper large-v3~\cite{radford2023whisper} for transcription and pyannote.audio~\cite{bredin2020pyannote} for speaker diarization, then manually verified the transcripts against the saved Zoom captions and screen recordings. The pilot and co-design sessions ran on earlier versions of the system and shaped its design; we exclude them from the findings.

We analyzed the session recordings and transcripts through thematic analysis, following Braun and Clarke's six-phase process~\cite{BraunClarke2006}. Two of the authors familiarized themselves with the full dataset, then coded independently at the utterance level, meeting regularly to compare codes and resolve discrepancies. They iteratively refined the codebook until it stabilized, grouping related excerpts into candidate themes and then into higher-level themes. Following McDonald et al.~\cite{mcdonald2019reliability}, we developed codes interpretively and resolved disagreements by consensus rather than computing inter-rater reliability. We also derived descriptive measures from the behavioral logs of all 14 sessions (e.g., dwell time per zone, revisits, and the order of touches and questions) and read them against the themes.

\section{Findings}
\label{sec:findings}

\subsection{RQ1: What each modality delivered}

\label{sec:f-modalities}

Participants took a place's \emph{where} from the non-verbal modalities and its \emph{what} from the verbal ones, and the agent gave both (Table~\ref{tab:modalities}). A place reached them in this order: the hum led the finger to it, the vibration gave its edge, the spoken name identified it, and a question told what it was for. They read one modality against another, checking a spoken direction by finger and asking about a place found by finger; where one modality did not reach a participant, another carried its information.

\begin{table}[tb]
\centering
\footnotesize
\renewcommand{\arraystretch}{1.2}
\caption{What each modality delivered, with the participants' own accounts in the last column. The \emph{where} is a place's position; the \emph{what} is its name and meaning.}
\begin{tabular}{@{}>{\raggedright\arraybackslash}p{0.13\linewidth} >{\raggedright\arraybackslash}p{0.30\linewidth} >{\raggedright\arraybackslash}p{0.44\linewidth}@{}}
\toprule
\multicolumn{1}{>{\centering\arraybackslash}p{0.13\linewidth}}{\textbf{modality}} & \multicolumn{1}{>{\centering\arraybackslash}p{0.30\linewidth}}{\textbf{what it delivered}} & \multicolumn{1}{>{\centering\arraybackslash}p{0.44\linewidth}}{\textbf{in the participants' words}} \\
\midrule
hum (audio) & the \emph{where} --- proximity: a place's closeness, and a direction to follow toward places not yet touched & \quot{it gets stronger and stronger as soon as you get near. It's like the radar on a car.}~(P13)\newline \quot{it starts humming, and once you get there it says [the center].}~(P13) \\
\midrule
vibration (haptic) & the \emph{where} --- boundary: the edge of a place, in or out, its size, and arrival & \quot{Because I've got this boundary --- from here to here. That's mapping in.}~(P04)\newline \quot{It's like a wall. So the vibration is inside only.}~(P14)\newline \quot{It gives you strong feedback that you are there.}~(P14) \\
\midrule
spoken name (audio) & the \emph{what} --- identity: the places present, each by its name & \quot{It's not telling me how to get from one to the other, but it is telling me what's there.}~(P07) \\
\midrule
agent (audio) & the \emph{where} --- direction: to a queried place, in the map's cardinal terms\newline the \emph{what} --- semantics: a place's kind, use and details, beyond its name & \quot{having the map and it spelling out each of the buildings, but then you're able to go deeper.}~(P02) \\
\bottomrule
\end{tabular}
\Description{What each modality delivered, with the participants' own accounts in the last column. The \emph{where} is a place's position; the \emph{what} is its name and meaning.}
\label{tab:modalities}
\end{table}

\textbf{A device new to most.} Few had used a trackpad, and P14 took it for part of the keyboard. Seven called the system easy. P02, told during onboarding that a question is put by holding the key: \quot{And wait and ask. Very clever. I love the simplicity of it.} P15 described the hour as a curve: \quot{At first, it was awkward because it's something brand new. And then the more I used it \ldots{} okay, that's up here~on~the~pad.}

\subsubsection{From approaching a place to asking about it}

\label{sec:f-approach}

\textbf{Searching by the hum.} Four searched with the hum and read nearness from it. P15 swept for one: \quot{I just start low, and start going up and sideways \ldots{} until I start hearing a hum. And then I follow the hum.} P13 and P14 read it the same way. P07 read how crowded the ground was from how much the hum varied as stages were added: \quot{It was very open area. As we added more buildings it (hum) stays on all the time.}

\textbf{The vibration as boundary.} For participants who felt it, the vibration marked the boundary and held while the finger was inside. P03: \quot{It's like, oh, I'm at the right spot.} Crossing a place gave its size, \quot{an idea of how big the space is} (P14). Following the boundary around gave a shape, which P04 said while still tracing: \quot{So far I'm touching something rectangular... Because I've got this boundary --- from here to here.} Nine of the ten participants who worked the practice square touched all four of its sides, and their fingers kept to the edges. That is contour following, what blindfolded observers do to judge an object's exact shape~\cite{lederman1987hand}.

\label{sec:f-differ}

\textbf{One modality standing in for another.} Where a participant did not perceive one modality, another carried complementary information. P12 heard the vibration and could not feel it: \quot{Yeah, I can hear it. The vibration somewhat... But I cannot feel it from the finger.} P02 and P15 heard it more than they felt it. A matching buzz plays with each vibration, so these three read boundaries by ear. P07 found the buzz \quot{annoying real fast} and still asked us to keep it for a blind reader unlike them, and P11 asked the same for a blind reader with little hearing.

\textbf{A name without a kind.} A name alone left participants without a target for their questions. The pad speaks a place's name on entry and leaves its kind to a question. Participants heard the name, found where the place was by touch, and still asked what it was for (P02, P04, P11). P11: \quot{What does '222' mean? Now I'm in 222.} P06 heard the Glass House named, took it for something to do with glass art, and asked what it was; the answer put it as a cafe or a place to get a snack. P04, asked afterwards what they had missed: \quot{Because I didn't know that beforehand, that's why I was asking more or less randomly.}

\textbf{Asking to the depth of a decision.} Participants went deeper, asking at the detail a decision needed. With the kind in hand they went further into a place: P03 asked about one restaurant's menu, then \quot{What are their store hours and what's their number?}, and about another whether it took reservations, keeping or dropping each place on the answer. Thirteen of the fourteen asked questions of this kind, running to hours, menus, prices, restrooms, and whether a summer camp takes visually impaired children.

\subsubsection{Reading one modality against another}

\label{sec:f-crosscheck}

\textbf{Checks in both directions.} Checks ran from an answer to the finger (8 of 14) and from the finger to a question (10 of 14), and we saw two kinds of check. On direction, P02, told a market lay east of them, swept east until the pad named it, and asked which place the answer was measured from, \quot{So when it said east, is it east of this Cat and Cloud?} On distance, P06 crossed the gap between the accessibility center and a physical therapy practice, called it \quot{about a block}, then asked the agent for it in feet. A right answer did not always deliver the place. P03, told four missing places lay to the northwest, swept over one without counting it, \quot{It's supposed to be up here, but it's not}, and afterwards could not say whose fault it was, \quot{I don't know if the AI was hiding it or not.}

\textbf{Where and what, when needed.} P04 knew the opposite from their phone assistant: \quot{Gemini tells you what's around you, but not where it is around you.} Every answer here begins by stating where the finger is (Section~\ref{sec:system-frontend}). When the question was a what-question, that position statement came before the answer, and P04 and P07 noticed. P07, with a finger on a building's outline: \quot{it would start describing the building as square and I'm going, I know.} P04, asking about business hours, was told they were in an open area: \quot{it didn't even answer the question.} P11 used the agent for what they had not already found by hand: \quot{I don't care what is here on the map because I can search for that with my finger...}

\subsection{RQ2: What participants worked out on the surface}

\label{sec:f-built}

Participants worked out where the map ended, how far apart its places were, and how much of it arrived at once. On that frame they did three things with a place. They discovered places they had not asked about and anchored them to places already on the map. Where their memory and the map disagreed, they revised one or the other. And they checked a spoken claim against what they had felt or been to.

\subsubsection{The frame: the rim, the scale, and how much arrives at once}

\label{sec:f-padtoreal}

\textbf{The rim and the scale.} Participants took the edge of the pad as the map's limit, and the gap they crossed between two places as its scale. P08 met the rim, \quot{If I keep going right I'm going to hit the table. There's no more of the pad.} For scale, three crossed the gap between two places and put it in a unit they walk in, a block (P04, P06, P12). P04: \quot{it lets me know how far I'm going to be walking or how long I should walk before I make a turn.} That unit made a number usable; left to choose, sighted and blind people alike judge distance by the walk rather than the straight line~\cite{rieser1980role}. P06: \quot{Some people don't really get 400 feet... A half a block explained a lot for me.}\looseness=-1

\textbf{Stages, and where to cut them.} Asked whether the map should arrive in stages or whole, five of the six asked kept the stages; P11 wanted everything at once. P09, whose session ran through a Spanish interpreter, described the stages as a place filling in: \quot{me va alzando las ubicaciones y las cosas que están alrededor mías como si lo estuviera mirando. Me sentí súper contento y libre de poder yo hacer las cosas yo solo} (it keeps raising up for me the locations and the things that are around me as if I were looking at it. I felt super happy and free to be able to do things by myself). Four named a basis for cutting the map other than the system's area and prominence (Section \ref{sec:system-frontend}): by the kind of place (P04), by price or how busy a street is (P13), by the unit a step moves in (P14), and by the trip, the full map when planning and two buildings when working inside one (P07).

\subsubsection{Discovering: meeting places they had not asked about, and anchoring them}

\label{sec:f-discovery}

\textbf{Places nobody had asked for.} Seven participants named a place they had not asked about. P04 works at the center on their map and had known five or ten of its fifty-five places, \quot{before I only knew of five --- literally only five, maybe 10,} and met a post office and a yoga studio a few doors from where they work. P07 said the same of a campus they would otherwise have walked: \quot{it's giving me a lot more around it. If I was going there just for O\&M, I probably would have wandered around campus but I probably would not have found all of those.}

\textbf{One place is not enough.} Most participants placed a first place by relating it to a second place or a street (10 of 14). P03, on a first stage holding one restaurant they already knew: \quot{there's only one place, it's like empty. It's like, oh, it's floating in nowhere.} P11 said the same of a map new to them. The second one was either a place they already knew or expected, often a corner (P04), or a street running alongside, brought in from outside the map. P06: \quot{it will actually show you right here which way Soquel Avenue travels.}

\begin{figure*}[t]
\centering
\includegraphics[width=\textwidth]{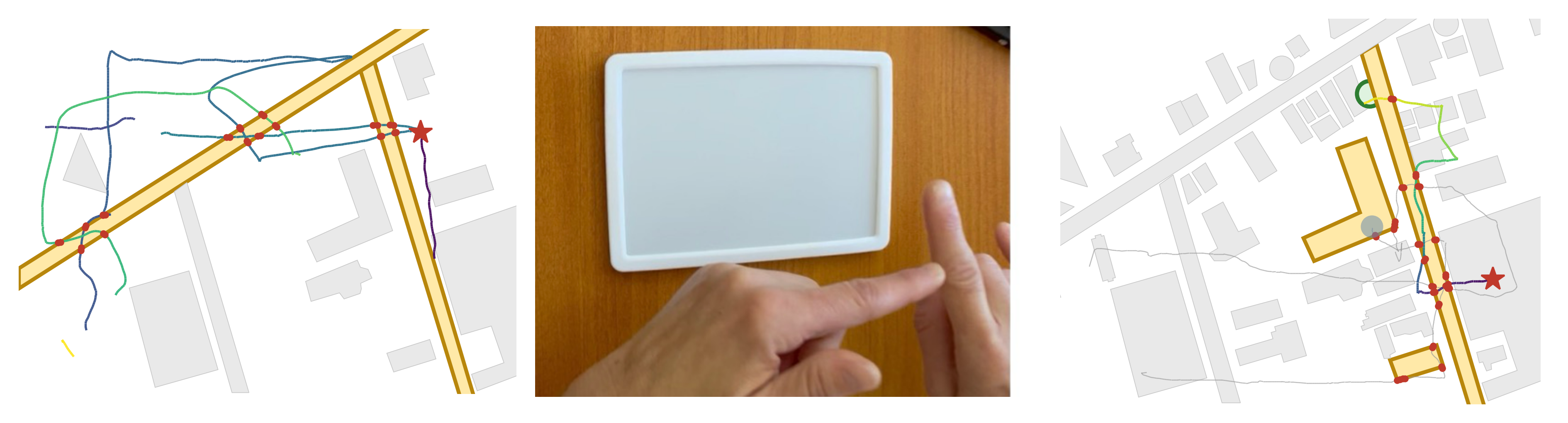}
\caption{Left, P06 finds that two streets cross: the finger runs Soquel Avenue, then Seabright Avenue, and the red samples, one per crossing of a street's edge, gather where the two meet; ``When you get here, Soquel Avenue is over here... it travels in a flat like this,'' then ``Seabright here kind of runs.'' Middle, immediately afterwards, both hands off the pad. For onboarding the study ran on a standalone Magic Trackpad and a standalone three-key keypad, both unmodified. Right, later, asked to find the bus stop: ``I'm just going to scan,'' and the finger sweeps out over the accessibility center and back (the thin gray line). From the parking lot (the gray-blue dot) P06 asks how far the stop is from the center and which way, naming the center rather than touching it. The answer places the stop to the north; P06 lifts off while it plays, lands beside the road, runs north past the crossing found earlier, and reaches the stop (outlined green): the colored line, dark early to bright late. In both panels the star marks where the stretch begins and gold the streets or places the episode turns on.}
\Description{Three panels. Left, a map with two gold street bands and a finger path in red samples gathering where the streets meet. Middle, a photograph of P06 with both hands raised, two fingers crossed in the air. Right, the same map with a gray search line, a star, a colored run north along the road, and a bus stop outlined in green.}
\label{fig:p006-crossing-gesture-photo}
\end{figure*}

\textbf{From anchor to relation.} With a second place named, participants said where the first sat relative to it (9 of 14). Streets were laid down first. P02 followed Soquel as the name repeated under the finger, felt the shape of its run, and said what the line was for: \quot{I can kind of go up and down from it to see what is along the street.} With Soquel settled, P06 ran a second name and said where the two would meet: \quot{All this is Seabright. And then if I go here somewhere, I'm going to be on Soquel Avenue.} The prediction held, and P06 lifted both hands off the pad and held two fingers across each other in the air beside it (Figure~\ref{fig:p006-crossing-gesture-photo}).

\subsubsection{Discovering: what they carried away}
\label{sec:f-carried}

\textbf{What they carried away.} At the end, a rough shape, a few names, or the run of the streets came back (7 of 14). The system had described P02's map as an hourglass; they gave it back as a butterfly: \quot{I'm thinking of kind of butterfly-shaped almost, where down in the center is kind of where the Cat and Cloud is.} P06 had already shown their own shape mid-session as the two crossed fingers of Figure~\ref{fig:p006-crossing-gesture-photo}, and P03 gave the plaza as a list of places. Asked which feature they liked most, P04 gave back the run of the streets: \quot{I thought it would be hard for me to build a mental map, but I was able to kind of draw that L-shape or square shape of the streets.} Asked afterwards, away from the pad, no participant gave a distance in feet or a precise outline.

P03 held a place in one of two forms, an area or a point, and moved between them. A new place was first explored as an area, then held as a point, a name at a rough position rather than an outline; a known place moved the other way. The plaza their family drives them to had been one point, \quot{I just know it's one place}; by the end they could name four businesses in it, \quot{in my mind there was only one spot... So now there are a lot of spots.} A Walgreens new to them went the other way, from an area under the finger to a point they measured from, \quot{Is HS Academy across the~street~from~Walgreens?}

\subsubsection{Revising: putting the map against what they already held}

\label{sec:f-against}

\textbf{Memory put to the map.} Ten participants read a place they knew against the map. P10 enumerated and verified a few places, then predicted one more from memory, \quot{I think across the street is the roller rink,} which the facilitator doubted was on the map; a minute later the pad named it, \quot{There's the roller rink.} P11 read a place they knew and found the map close to their memory of it, then asked whether that made the map reliable for a place they did not know: \quot{can I count on this for a place I don't know? That's my question.} P04 walks Soquel and took its run from it: \quot{I don't always feel how Soquel is moving south to north. But here it helped me understand that.}

\textbf{Two routes from home, no line between them.} Two places a few minutes apart had been two separate trips from home, and had stayed that way. P06 commutes from another town to the accessibility center and uses a pharmacy a five-minute walk from it. Crossing a row of places as the pad spoke each name, they stopped on the pharmacy: \quot{I've never been there from here though. But I've been from my house to get this walker. But I didn't know it was this close to [the center].\ldots{} See, this map --- I learned something today.} They had held it in another town and reached it by going home first: \quot{I would go all the way [home] and \ldots{} come all the way back to the same place I just left.} The place was not new to them; its position was. \quot{And no one tells you. It could be that way forever.}

\subsubsection{Checking: when an answer disagreed with what they had felt or been to}

\label{sec:f-conflict}

\textbf{An answer put to memory.} Five participants met an answer against what they had felt or been to, and four of them kept their own. We saw three kinds of disagreement. \emph{Distance and direction}: P03 set the agent's distance against the one they had felt between three shops, \quot{So why is it further from Walgreens when it's closer?}; P08 misread a correct direction and trusted their walks instead, \quot{I've gone there many times.} \emph{Attribute}: P14 had eaten at the restaurant and overruled the answer, \quot{Bechamel is not vegan. I've been to the restaurant.} \emph{Existence}: the agent missed a store P13 had been to and claimed it was online only. Asked afterwards, P13 kept what they knew, \quot{mainly because I know that store is there and I've been to it}. P07, the fifth, doubted their memory instead of the answer, \quot{it could have been this.}

\textbf{Grounds for believing.} Asked why they believed an answer, participants each gave their own grounds (7 of 14). P07 called an answer detailed but wanted to look it up, and P15 said the question never arose, \quot{I didn't even consider if I trusted it or not.} Asked at the end, three said they trusted the agent as a whole and had seen no mistake (P02, P06, P09). P13 named the store error as the one thing that had surprised them, counted it as ordinary, \quot{Google Maps glitches sometimes, or Siri does too}, and kept their trust: catching a claim and trusting the machine are separate judgements.

\subsection{RQ3: How that understanding enters life}

\label{sec:f-life}

Participants said what would change for them after reading the map. What they would keep was the surroundings of a place: to know where they are, to have somewhere to go when a plan fails, and to tell others. A place they had found became a stop on a trip they make or a trip of its own. And they placed the reading at home before a trip, beside the tools that carry the walk.

\subsubsection{Knowing what is around, for themselves and for others}

\label{sec:f-surround}

\textbf{Surroundings rather than a route.} Three participants came for what is around a destination rather than a route to it (P04, P06, P08); P11 kept to the route and the app they trust. P06: \quot{Because that was my whole point in the beginning --- having surrounding awareness.} P04, P06 and P08 also said that what is around them is how they know where they are. P08 used landmarks to check their position: \quot{They just help me know where I'm at, that I'm not lost.}

\textbf{A fallback when a plan fails.} Two named the places around a destination as their fallback when a plan fails (P06, P08). P08 named them while still on the map: \quot{there's a bus stop here but if I miss the bus stop there's a restaurant or there's Ace Hardware I can go to.} P06 named a failure of another kind: \quot{What if you're feeling sick? At least you know 'okay, I know what's around me. I need to go sit down.'}\looseness=-1

\textbf{Something to pass on.} For P04, knowing what was around became something to pass on. P04 takes the calls at the center from people who come in from out of town and stay for hours: \quot{they don't know what's around them. So now I'm learning --- where to eat, where to take a break, maybe have fun, where to drop off mail at USPS.} They could give a caller directions that ran past the places the map had named: \quot{you just walk out of [the center], turn to that direction, pass yoga --- which I never knew there was one --- the Bagelry restaurant.\ldots{} I can tell someone sighted how to get somewhere now.}

\subsubsection{From a discovery to a trip}

\label{sec:f-plan}

\textbf{A found place changes a trip.} Four participants added a stop to a trip they already make or made a trip of its own. P06 had held the pharmacy of Section \ref{sec:f-against} in another town: \quot{now that I know [the pharmacy] is right there and it's on my way home, I just stop there instead of going all the way home and coming all the way back.} P08 likewise had a use for a rug cleaner new to them: \quot{I have some rugs at my house I need cleaned}. For the other two the discovery itself became the reason to go. P11 had tried the same place in the map application they use: \quot{This is the first time I know there is a statue for Nikola Tesla. And I'm interested to take a picture there.} P07 asked the agent about the exhibits at several museums on the map and left with more places than they expected: \quot{I didn't think there would be that many things on the one map \ldots{} that I was like, oh yeah, I need to go.\ldots{} I should look them up on Google Maps \ldots{} and start plotting out where I want to go.} P06 also planned for someone else. They recalled a trip with a seven-year-old granddaughter to an appointment at a transit center, and said a map like this would show places to take her to eat and let them plan ahead: \quot{it makes me feel safe with her.}

\subsubsection{Placing the map in their daily life}

\label{sec:f-place}

\textbf{At home, before the trip.} Participants placed the reading at home, before the trip. We introduced every session that way, and three named the use in their own words first (P04, P06, P14); the rest agreed once we put it to them. P14: \quot{Because I don't think you can use this on the go. This exploration is done before you leave your house.} Two of the three also wanted the same signals while walking. P06 wanted the hum in headphones, and P14 wanted the agent in glasses.

\textbf{Alongside the tools they use.} Four participants placed this map alongside the tools they already use. Asked to compare, P11 declined, \quot{I don't prefer to compare it. I mean, it's complementary,} and P10 called them \quot{two different things}. P11 wanted the start from this map and the walk from another, \quot{If I arrive at the starting point alone I can continue with another map provider --- Google Maps or Apple Maps,} and did so with the statue, \quot{Hey Gemini, I am in the entrance of [the center]. I need to get to the Nikola Tesla statue. Guide me please.} P07 left with places to look up in a route planner. P04 set the map against the assistant used standing in the street, \quot{you're only going to see what's immediately in front of you.} P14 asked for the two to connect, \quot{apps are independent. They're like islands.}

\textbf{Requirements for a street version.} P14 set the requirements a version on the street would have to meet, keeping the table version for home. It would have to run on a phone or a tablet; go without a button, since \quot{You'd have to let go of your cane to press the button and then grab your cane again}, and without gestures in the air, since \quot{you'd look kind of strange walking down the street making these hand movements}; not be held up in front, since \quot{some blind people have had their iPhone stolen that way}; and be heard without masking the street, \quot{earphones where you can hear it but still hear what's around you safely.} Smart glasses would meet most of these, \quot{But I'm worried about privacy. I'm worried about having to pay a subscription.}

\section{Discussion}
\label{sec:discussion}

Accessible maps make a layout explorable by hand or convey a place in speech, and a blind traveler's picture of a place is assembled from whichever of these they have. Fourteen blind and low-vision participants read a map that does both: a place they chose, on a laptop trackpad where each point of the pad is a fixed place on the map, with an agent that answers about the place under the finger. Because a place kept its position, participants could return to it and locate the next place relative to it. Because the whole layout lay under the hand, they chose the order, the depth and the amount of their reading. On that surface they discovered places nobody had named, revised what they held and what the map held, and checked spoken claims by finger. They placed this reading at home, before a trip, and asked for a version the street would allow.

\subsection{Relative tracking sends a movement, absolute mapping sends a place}

A trackpad normally reports how far the finger moved. Our mapping makes it report where the finger is, and that point is a fixed place on the map (Table~\ref{tab:relative-absolute}). For a reader the difference is whether a place can be found again. With a coordinate a reader remembers one position, returns to it, and ties a place met later to the one they started from. With a sum of movements there is no position to remember, so a place found once is gone on the next move. In the first pilot the co-designer could not hold a picture from a relative pointer and asked for a reference that would keep its place, and on the absolute mapping ten of the fourteen participants set a first place against a second place or a street (Section~\ref{sec:f-discovery}).

\begin{table}[tb]
\centering
\footnotesize
\renewcommand{\arraystretch}{1.2}
\begin{tabular}{@{}>{\raggedright\arraybackslash}p{0.14\linewidth} >{\raggedright\arraybackslash}p{0.37\linewidth} >{\raggedright\arraybackslash}p{0.40\linewidth}@{}}
\toprule
\textbf{aspect} & \textbf{relative tracking (the default)} & \textbf{absolute mapping (here)} \\
\midrule
what the pad sends & a movement; the position is summed from the movements so far & a coordinate; the place is where the finger rests \\
what a slide means & a direction of travel, started from wherever the finger lands & a change of place; lifting and landing elsewhere is elsewhere \\
edge and stages & the map has no edge, so one place can be met at the top left and again at the bottom right; a further stage has no fixed position against the last & the rim is the edge; a further stage is the same map with more on it \\
\bottomrule
\end{tabular}
\caption{What the pad sends, what a slide means, and where the map ends, under relative and under absolute tracking.}
\Description{What the pad sends, what a slide means, and where the map ends, under relative and under absolute tracking.}
\label{tab:relative-absolute}
\end{table}

A spoken route is a sequence of moves and a spoken set of places is a list. Both give one relation at a time, as relative tracking does, and the listener must assemble the layout. P04 writes a route as a string of turns, \quot{So I just write 3L 2R 3L}, and P02 said a blind reader holds a great deal with nothing to write on. On the surface the whole layout was present at once, and a relation was there to be read where the finger already was; one participant who had settled a street named the street that would cross it before meeting it. Haptic and auditory information is taken in one part at a time, while vision takes in a scene at once~\cite{ottink2022cognitive}. Our surface sits between: the hand reads one part at a time, but the whole layout stays in place under it. We read this through external cognition. A representation indexed by location lets a reader take a relation off it with no sequence to assemble~\cite{larkin1987diagram,scaife1996external}, and part of the remembering runs through the surface~\cite{hollan2000distributed}. Those accounts were written for diagrams read by eye; our results extend them to a surface read by hand, and mark where the extension stops. Away from the table the reader carries the layout: P03, reciting afterwards, reached two places they could no longer name.

Designers who put a map on a touch surface should settle three questions: what a touch tells the reader, how the reader gets back to a place they have found, and what each movement \mbox{of~the~finger~means.}

\subsection{Participants set the order, the depth and the amount of the reading}

Participants chose where to go next, how much to ask of one place, and how much of the map to have at once (Sections~\ref{sec:f-discovery} and~\ref{sec:f-padtoreal}). They moved between places in the order they wanted and went back to compare, asked further about a place, and took a further stage as the same place with more on it. The reader set three things, the order, the depth, and the amount; a spoken account lets a listener set none of them, since it comes in the order of its sentences. The same map heard as speech would come in one order, and P04 described the cost: a map application read twenty sponsored results before one nearby. Paper fixes the amount, as speech fixes the order. Participants asked for four ways to filter the map; the data are already on it, but on paper each filter would need its own printed sheet. A prior touch surface lets the reader ask for detail on request~\cite{zhao2024tada}. Our readers also set the order and the amount, and all three changed while the finger stayed on the place. Even our own staged order was a default the reader left: it held for a first reading, P11 wanted the whole map from the start, and four participants proposed a cut of their own. Designers of such a map should support the reader in setting the order, the detail and the density of a reading, since readers differ in how much of a map they take at once and in what they come to it for.

\subsection{Participants discovered, revised and checked a picture nobody else sees}

A blind traveler's picture of a neighborhood is assembled from their own routes and from what others have said~\cite{williams2013pray}, and it stays in their head. We call it a picture nobody else sees. A sighted traveler's picture is corrected by the street and the map without anyone intending it; a blind traveler's is not. A gap in it, an error in it, and a claim that contradicts it all pass unnoticed. On the pad each one showed.\looseness=-1

A gap in the picture goes unnoticed. P04 takes calls at the center asking what is around, and said the staff can barely answer that themselves. On the pad, places arrived under the finger that no one had asked about, several on ground the participant already traveled (Section~\ref{sec:f-discovery}). Routes and asking only fill a gap the traveler already knows about. On the pad the place arrived without a question. A place met this way started the next question and gave the reader something to measure the answer against. We see the pad supplying reference points a walk had not~\cite{couclelis1987anchor}; P11 described the same growth on foot, a landmark every hundred feet, linked on later walks. Evaluations of pre-travel access score the walk that follows, on questions the evaluator posed~\cite{espinosa1998comparing,lahav2008haptic}; such evaluations have no way to score a place the reader did not ask about.

An error in the picture goes uncorrected. Without the street or the map to correct it, a correction waits until the reader asks or a trip goes wrong. P06 believed the pharmacy was in another town and reached it by going home first. Two places learned on separate trips had never been placed against each other~\cite{montello1998framework}, until the map showed both at once (Section~\ref{sec:f-against}). Our results also showed the correction running the other way: a participant who had been to a place overruled the agent about what it served (Section~\ref{sec:f-conflict}). This extends the practice of testing a machine on what one already knows~\cite{alharbi2024misfitting}: here the reader tested the map and disputed it on the same map in the same sitting, so the reader's picture and the map's picture corrected each other.

A claim against the picture goes unchecked. A blind listener has been found to side with the claim~\cite{macleod2017captions}, and the checks open to them bring in another person, device or trip. On the surface a check took one reach. Participants asked what a spoken direction was measured from and went to see, and predicted where a place would be and put a finger there (Sections~\ref{sec:f-crosscheck} and \ref{sec:f-against}). A reader could check an answer only where the answer pointed to a spot on the surface. An answer grounded at the finger could be tried against the surface, as prior touch maps do by design~\cite{mapio2025,zhao2024tada}; an answer about what a place serves could not, and for such claims participants named checks of their own, a number to call (P03), a second source (P07), or what they already knew (P13). We read this as the limit of the surface as a check: a wrong claim about where a place lies stops at the finger, and a wrong claim about what it serves enters the picture the reader carries away. The more questions an agent answers, the more of its answers cannot be checked by touch, and it does not mark which ones those are.

Designers of a map a blind reader consults alone should make room for a discovery the reader did not ask for, for the map revising the reader's picture and the reader revising the map, and for a spatial check the reader can make against each claim.

\subsection{Participants placed the reading at home and wanted it on the street}

Participants placed this reading at home, before a trip, and left the walk to the tools they already carry (Section~\ref{sec:f-place}); four of them left with a stop added to a trip they make, or a trip of its own (Section~\ref{sec:f-plan}). They set it against what they had been offered before: a refreshable tactile display costing thousands of dollars, and devices that reach a reader through a rehabilitation counselor or an institution. P11 raised the price, was told the map runs on a laptop's own trackpad, and answered \quot{I can have it on my computer.\ldots{} the product itself is just software.} That answer rests on two facts. The hardware is the trackpad and actuator of an everyday laptop. The map is data, generated for a place and a reader in minutes from open map data, so what a map covers follows the data; P04 set that against the printed maps they had used, \quot{I like that it's dynamic --- not a static map where you constantly have to print something out.}%

Our results also showed the same participants asking for a version on the street, on the street's terms. On the street the reader loses what the table provided: a free second hand and a fixed edge. At the table a reader has two hands, one of them on the edge; on the street one hand holds the cane and hearing already carries the walk, and P14 listed what a street version would have to do without: a button, a gesture in the air, a phone held up, an earphone that masks the street. The body sets terms too. Some participants read the boundary by ear where they could not feel it (Section~\ref{sec:f-differ}), which is what universal design asks for when it puts essential information in more than one mode~\cite{connell1997principles}; the vibro-audio studies this work builds on paired vibration with audio and did not report a participant who felt nothing~\cite{giudice2012learning,palani2022comparing,giudice2020cognitive}. Several participants liked the three physical keys because they could feel and press them, P12 still carries a keyed GPS device because a flat screen gives their fingers nothing, and P14 asked for the hum's frequency to be the reader's to set. When in a trip the map is read decides what it runs on. In a pocket the map loses the edge and the second hand, and keys, gestures and speech do not replace them.

To serve its part of a trip, a pre-travel map of this kind could run on a device the reader already owns, carry each signal in more than one modality, and hand its reading on to the tools \mbox{that~carry~the~walk.}

\section{Conclusion}
BLV travelers can learn about named places and obtain route information through accessible maps, but few systems support discovering places they did not know existed. Building spatial understanding is an essential step before traveling to an unfamiliar place. To this end, we developed TouchingSpace, a trackpad-based audio-haptic map coupled with a conversational agent. The system retrieves geographic data for a place and maps it to fixed positions on a trackpad. Users receive audio and haptic feedback during exploration, and the conversational agent answers open-ended questions. We conducted a user study with fourteen BLV participants who freely explored a personally relevant place using our system. We found that participants used sound and vibration to locate places and trace their boundaries, while spoken names identified them and the agent related them to others. On the bounded surface, participants discovered places they had not asked about, revised prior beliefs, and checked spatial relations. They expected the resulting surrounding awareness to support future travel and complement the navigation tools they already use. These results suggest that accessible AI maps should pair conversational depth with a bounded spatial representation under the reader's control so that discovery and verification can occur on the same surface.

This study has several limitations. First, TouchingSpace has two potential sources of error: the generated map may inherit inaccurate labels or boundaries from geographic data, and the agent may answer complex questions incorrectly. Future work should develop algorithms and validation pipelines for higher accuracy. Second, participants proposed alternative ways to filter map content beyond the four stages implemented in TouchingSpace. Future versions should give readers more control over which places are shown and at what density. Finally, this study provides a starting point for understanding self-directed, pre-travel map exploration. Longitudinal studies with broader samples can extend this work by examining how spatial understanding persists and complements navigation tools during real trips.

\bibliographystyle{ACM-Reference-Format}
\bibliography{reference}

@article{brock2015interactivity,
  title = {Interactivity Improves Usability of Geographic Maps for Visually Impaired People},
  author = {Brock, Anke M. and Truillet, Philippe and Oriola, Bernard and Picard, Delphine and Jouffrais, Christophe},
  journal = {Human--Computer Interaction},
  volume = {30},
  number = {2},
  pages = {156--194},
  year = {2015},
  doi = {10.1080/07370024.2014.924412}
}

@incollection{ducasse2018tangible,
  title = {Accessible Interactive Maps for Visually Impaired Users},
  author = {Ducasse, Julie and Brock, Anke M. and Jouffrais, Christophe},
  booktitle = {Mobility of Visually Impaired People: Fundamentals and ICT Assistive Technologies},
  editor = {Pissaloux, Edwige and Vel{\'a}zquez, Ramiro},
  pages = {537--584},
  year = {2018},
  publisher = {Springer},
  address = {Cham},
  doi = {10.1007/978-3-319-54446-5_17}
}

@inproceedings{holloway2018accessible,
  title = {Accessible Maps for the Blind: Comparing 3D Printed Models with Tactile Graphics},
  author = {Holloway, Leona and Marriott, Kim and Butler, Matthew},
  booktitle = {Proceedings of the 2018 CHI Conference on Human Factors in Computing Systems},
  pages = {1--13},
  year = {2018},
  publisher = {ACM},
  doi = {10.1145/3173574.3173772}
}

@article{ottink2022cognitive,
  title = {Cognitive Map Formation Supported by Auditory, Haptic, and Multimodal Information in Persons with Blindness},
  author = {Ottink, Loes and Buimer, Hendrik and van Raalte, Bram and Doeller, Christian F. and van der Geest, Thea M. and van Wezel, Richard J. A.},
  journal = {Neuroscience \& Biobehavioral Reviews},
  volume = {140},
  pages = {104797},
  year = {2022},
  doi = {10.1016/j.neubiorev.2022.104797}
}

@inproceedings{reinders2020heymodel,
  title = {``Hey Model!'' -- Natural User Interactions and Agency in Accessible Interactive 3D Models},
  author = {Reinders, Samuel and Butler, Matthew and Marriott, Kim},
  booktitle = {Proceedings of the 2020 CHI Conference on Human Factors in Computing Systems},
  pages = {1--13},
  year = {2020},
  publisher = {ACM},
  doi = {10.1145/3313831.3376145}
}

@inproceedings{geovisa11y2026,
  title = {GeoVisA11y: An AI-based Geovisualization Question-Answering System for Screen-Reader Users},
  author = {Li, Chu and Pang, Rock Yuren and Chheda-Kothary, Arnavi and Sharif, Ather and Assalif, Henok and Heer, Jeffrey and Froehlich, Jon E.},
  booktitle = {Proceedings of the 2026 CHI Conference on Human Factors in Computing Systems},
  year = {2026},
  publisher = {ACM},
  note = {arXiv:2603.07446},
  doi       = {10.1145/3772318.3790334},
  numpages= {17},
}

@article{lee2004trust,
  title = {Trust in Automation: Designing for Appropriate Reliance},
  author = {Lee, John D. and See, Katrina A.},
  journal = {Human Factors},
  volume = {46},
  number = {1},
  pages = {50--80},
  year = {2004},
  doi = {10.1518/hfes.46.1.50_30392}
}

@article{hoff2015trust,
  title = {Trust in Automation: Integrating Empirical Evidence on Factors That Influence Trust},
  author = {Hoff, Kevin Anthony and Bashir, Masooda},
  journal = {Human Factors},
  volume = {57},
  number = {3},
  pages = {407--434},
  year = {2015},
  doi = {10.1177/0018720814547570}
}

@inproceedings{alharbi2024misfitting,
  title = {Misfitting With AI: How Blind People Verify and Contest AI Errors},
  author = {Alharbi, Rahaf and Lor, Pa and Herskovitz, Jaylin and Schoenebeck, Sarita and Brewer, Robin},
  booktitle = {Proceedings of the 26th International ACM SIGACCESS Conference on Computers and Accessibility (ASSETS '24)},
  year = {2024},
  doi = {10.1145/3663548.3675659},
  numpages= {17},
}

@inproceedings{cavazosquero2019jido,
  title = {Jido: A Conversational Tactile Map for Blind People},
  author = {Cavazos Quero, Luis and Iranzo Bartolom\'{e}, Jorge David and Lee, Dongmyeong and Lee, Yerin and Lee, Sangwon and Cho, Jundong},
  booktitle = {Proceedings of the 21st International ACM SIGACCESS Conference on Computers and Accessibility (ASSETS '19)},
  year = {2019},
  pages = {682--684},
  doi = {10.1145/3308561.3354600}
}

@inproceedings{scenescout2026,
  title = {SceneScout: Towards AI Agent-driven Access to Street View Imagery for Blind Users},
  author = {Jain, Gaurav and Findlater, Leah and Gleason, Cole},
  booktitle = {Proceedings of the 2026 CHI Conference on Human Factors in Computing Systems (CHI '26)},
  year = {2026},
  publisher = {ACM},
  doi = {10.1145/3772318.3790449},
  note = {arXiv:2504.09227},
  numpages= {22},
}

@inproceedings{adnin2024king,
  title = {``I look at it as the king of knowledge'': How Blind People Use and Understand Generative AI Tools},
  author = {Adnin, Rudaiba and Das, Maitraye},
  booktitle = {Proceedings of the 26th International ACM SIGACCESS Conference on Computers and Accessibility (ASSETS '24)},
  year = {2024},
  doi = {10.1145/3663548.3675631},
  numpages= {14},
}

@inproceedings{sakib2026xai,
  title = {Explainable AI for Blind and Low-Vision Users: Navigating Trust, Modality, and Interpretability in the Agentic Era},
  author = {Sakib, Abu Noman Md and Dey, Protik and Zhang, Zijie and Akter, Taslima},
  booktitle = {CHI 2026 Workshop on Human-Centered Explainable AI (HCXAI)},
  year = {2026},
  note = {arXiv:2604.00187}
}

@article{schinazi2016navigation,
  title = {Spatial navigation by congenitally blind individuals},
  author = {Schinazi, Victor R. and Thrash, Tyler and Chebat, Daniel-Robert},
  journal = {WIREs Cognitive Science},
  volume = {7},
  number = {1},
  pages = {37--58},
  year = {2016},
  doi = {10.1002/wcs.1375}
}

@article{mapio2025,
  title = {MapIO: A Gestural and Conversational Interface for Tactile Maps},
  author = {Manzoni, Matteo and Mascetti, Sergio and Ahmetovic, Dragan and Crabb, Ryan and Coughlan, James M.},
  journal = {IEEE Access},
  volume = {13},
  year = {2025},
  doi = {10.1109/ACCESS.2025.3566286},
  pages   = {84038--84056},
}

@inproceedings{froehlich2025streetviewai,
  title = {StreetReaderAI: Making Street View Accessible Using Context-Aware Multimodal AI},
  author = {Froehlich, Jon E. and Fiannaca, Alex and Jaber, Nimer and Tsaran, Victor and Kane, Shaun},
  booktitle = {Proceedings of the 38th Annual ACM Symposium on User Interface Software and Technology (UIST '25)},
  year = {2025},
  publisher = {ACM},
  doi = {10.1145/3746059.3747756},
  numpages= {22},
}

@article{palani2022comparing,
  title = {Comparing Map Learning between Touchscreen-Based Visual and Haptic Displays: A Behavioral Evaluation with Blind and Sighted Users},
  author = {Palani, Hari Prasath and Fink, Paul D. S. and Giudice, Nicholas A.},
  journal = {Multimodal Technologies and Interaction},
  volume = {6},
  number = {1},
  pages = {1},
  year = {2022},
  doi = {10.3390/mti6010001}
}

@article{couclelis1987anchor,
  title = {Exploring the anchor-point hypothesis of spatial cognition},
  author = {Couclelis, Helen and Golledge, Reginald G. and Gale, Nathan and Tobler, Waldo},
  journal = {Journal of Environmental Psychology},
  volume = {7},
  number = {2},
  pages = {99--122},
  year = {1987},
  doi       = {10.1016/s0272-4944(87)80020-8}
}

@techreport{connell1997principles,
  title = {The Principles of Universal Design, Version 2.0},
  author = {Connell, Bettye Rose and Jones, Mike and Mace, Ron and Mueller, Jim and Mullick, Abir and Ostroff, Elaine and Sanford, Jon and Steinfeld, Ed and Story, Molly and Vanderheiden, Gregg},
  institution = {Center for Universal Design, North Carolina State University},
  year = {1997}
}

@article{jiang2026electrotactile,
  title = {LLM-powered assistant with electrotactile feedback to assist blind and low vision people with maps and routes preview},
  author = {Jiang, Chutian and Fan, Yinan and Xie, Junan and Kuang, Emily and Zhang, Kaihao and Fan, Mingming},
  journal = {International Journal of Human-Computer Studies},
  volume = {207},
  pages = {103682},
  year = {2026},
  doi       = {10.1016/j.ijhcs.2025.103682}
}

@misc{microsoft2018soundscape,
  title = {Project Soundscape: Maps Delivered in 3D Sound},
  author = {{Microsoft Research}},
  howpublished = {\url{https://www.microsoft.com/en-us/research/project/soundscape-maps-delivered-in-3d-sound/}},
  year = {2017},
  note = {Project page; established 1 January 2017, the project has concluded and the application was released as open source in June 2023}
}

@article{thorndyke1982maps,
  author  = {Thorndyke, Perry W. and Hayes-Roth, Barbara},
  title   = {Differences in spatial knowledge acquired from maps and navigation},
  journal = {Cognitive Psychology},
  volume  = {14},
  number  = {4},
  pages   = {560--589},
  year    = {1982},
  month   = oct,
  doi     = {10.1016/0010-0285(82)90019-6},
  publisher = {Elsevier}
}

@inproceedings{kane2011access,
  author = {Kane, Shaun K. and Morris, Meredith Ringel and Perkins, Annuska Z. and Wigdor, Daniel and Ladner, Richard E. and Wobbrock, Jacob O.},
  title = {Access Overlays: Improving Non-Visual Access to Large Touch Screens for Blind Users},
  booktitle = {Proceedings of the 24th Annual ACM Symposium on User Interface Software and Technology (UIST '11)},
  year = {2011},
  pages = {273--282},
  publisher = {ACM},
  address = {New York, NY, USA},
  doi = {10.1145/2047196.2047232}
}

@inproceedings{macleod2017captions,
  author    = {Haley MacLeod and Cynthia L. Bennett and Meredith Ringel Morris and Edward Cutrell},
  title     = {Understanding Blind People's Experiences with Computer-Generated Captions of Social Media Images},
  booktitle = {Proceedings of the 2017 {CHI} Conference on Human Factors in Computing Systems},
  series    = {CHI '17},
  pages     = {5988--5999},
  publisher = {ACM},
  address   = {Denver, CO, USA},
  year      = {2017},
  doi       = {10.1145/3025453.3025814},
  url       = {https://doi.org/10.1145/3025453.3025814}
}

@article{thinusblanc1997representation,
  author  = {Thinus-Blanc, Catherine and Gaunet, Florence},
  title   = {Representation of space in blind persons: Vision as a spatial sense?},
  journal = {Psychological Bulletin},
  year    = {1997},
  volume  = {121},
  number  = {1},
  pages   = {20--42},
  doi     = {10.1037/0033-2909.121.1.20},
  publisher = {American Psychological Association}
}

@inproceedings{bigham2010vizwiz,
  author    = {Jeffrey P. Bigham and Chandrika Jayant and Hanjie Ji and Greg Little and Andrew Miller and Robert C. Miller and Robin Miller and Aubrey Tatarowicz and Brandyn White and Samuel White and Tom Yeh},
  title     = {{VizWiz}: nearly real-time answers to visual questions},
  booktitle = {Proceedings of the 23rd Annual {ACM} Symposium on User Interface Software and Technology},
  series    = {UIST '10},
  address   = {New York, NY, USA},
  pages     = {333--342},
  publisher = {{ACM}},
  year      = {2010},
  doi       = {10.1145/1866029.1866080},
  url       = {https://doi.org/10.1145/1866029.1866080}
}

@inproceedings{williams2013pray,
  author    = {Williams, Michele A. and Hurst, Amy and Kane, Shaun K.},
  title     = {"Pray before you step out": describing personal and situational blind navigation behaviors},
  booktitle = {Proceedings of the 15th International ACM SIGACCESS Conference on Computers and Accessibility},
  series    = {ASSETS '13},
  year      = {2013},
  publisher = {Association for Computing Machinery},
  address   = {New York, NY, USA},
  location  = {Bellevue, Washington, USA},
  articleno = {28},
  pages     = {28:1--28:8},
  numpages  = {8},
  doi       = {10.1145/2513383.2513449},
  url       = {https://doi.org/10.1145/2513383.2513449}
}

@article{bucinca2021trust,
  author       = {Bu\c{c}inca, Zana and Malaya, Maja Barbara and Gajos, Krzysztof Z.},
  title        = {To Trust or to Think: Cognitive Forcing Functions Can Reduce Overreliance on AI in AI-assisted Decision-making},
  journal      = {Proceedings of the ACM on Human-Computer Interaction},
  volume       = {5},
  number       = {CSCW1},
  articleno    = {188},
  numpages     = {21},
  year         = {2021},
  month        = apr,
  publisher    = {Association for Computing Machinery},
  address      = {New York, NY, USA},
  doi          = {10.1145/3449287}
}

@inproceedings{abdolrahmani2018siri,
  author    = {Abdolrahmani, Ali and Kuber, Ravi and Branham, Stacy M.},
  title     = {{``Siri Talks at You'': An Empirical Investigation of Voice-Activated Personal Assistant (VAPA) Usage by Individuals Who Are Blind}},
  booktitle = {Proceedings of the 20th International ACM SIGACCESS Conference on Computers and Accessibility (ASSETS '18)},
  year      = {2018},
  pages     = {249--258},
  publisher = {ACM},
  address   = {New York, NY, USA},
  location  = {Galway, Ireland},
  doi       = {10.1145/3234695.3236344}
}

@inproceedings{radford2023whisper,
  title     = {Robust Speech Recognition via Large-Scale Weak Supervision},
  author    = {Radford, Alec and Kim, Jong Wook and Xu, Tao and Brockman, Greg and McLeavey, Christine and Sutskever, Ilya},
  booktitle = {Proceedings of the 40th International Conference on Machine Learning},
  series    = {Proceedings of Machine Learning Research},
  volume    = {202},
  pages     = {28492--28518},
  year      = {2023},
  publisher = {PMLR},
  url       = {https://proceedings.mlr.press/v202/radford23a.html}
}

@inproceedings{bredin2020pyannote,
  title     = {{pyannote.audio}: neural building blocks for speaker diarization},
  author    = {Bredin, Herv\'{e} and Yin, Ruiqing and Coria, Juan Manuel and Gelly, Gregory and Korshunov, Pavel and Lavechin, Marvin and Fustes, Diego and Titeux, Hadrien and Bouaziz, Wassim and Gill, Marie-Philippe},
  booktitle = {ICASSP 2020 - 2020 IEEE International Conference on Acoustics, Speech and Signal Processing (ICASSP)},
  year      = {2020},
  month     = may,
  pages     = {7124--7128},
  publisher = {IEEE},
  doi       = {10.1109/ICASSP40776.2020.9052974}
}

@article{mcdonald2019reliability,
  author       = {McDonald, Nora and Schoenebeck, Sarita and Forte, Andrea},
  title        = {Reliability and Inter-rater Reliability in Qualitative Research: Norms and Guidelines for CSCW and HCI Practice},
  journal      = {Proceedings of the ACM on Human-Computer Interaction},
  year         = {2019},
  volume       = {3},
  number       = {CSCW},
  articleno    = {72},
  numpages     = {23},
  pages        = {1--23},
  month        = nov,
  doi          = {10.1145/3359174},
  publisher    = {Association for Computing Machinery},
  address      = {New York, NY, USA}
}

@misc{gemini25,
  author        = {{Gemini Team, Google}},
  title         = {Gemini 2.5: Pushing the Frontier with Advanced Reasoning, Multimodality, Long Context, and Next Generation Agentic Capabilities},
  year          = {2025},
  eprint        = {2507.06261},
  archivePrefix = {arXiv},
  primaryClass  = {cs.CL},
  url           = {https://arxiv.org/abs/2507.06261}
}

@inproceedings{su2010timbremap,
  author    = {Su, Jing and Rosenzweig, Alyssa and Goel, Ashvin and de Lara, Eyal and Truong, Khai N.},
  title     = {Timbremap: enabling the visually-impaired to use maps on touch-enabled devices},
  booktitle = {Proceedings of the 12th International Conference on Human-Computer Interaction with Mobile Devices and Services (MobileHCI '10)},
  year      = {2010},
  pages     = {17--26},
  publisher = {ACM},
  address   = {New York, NY, USA},
  doi       = {10.1145/1851600.1851606}
}

@inproceedings{giudice2012learning,
  author    = {Giudice, Nicholas A. and Palani, Hari Prasath and Brenner, Eric and Kramer, Kevin M.},
  title     = {Learning non-visual graphical information using a touch-based vibro-audio interface},
  booktitle = {Proceedings of the 14th International ACM SIGACCESS Conference on Computers and Accessibility (ASSETS '12)},
  year      = {2012},
  pages     = {103--110},
  address   = {Boulder, Colorado, USA},
  publisher = {ACM},
  doi       = {10.1145/2384916.2384935},
  isbn      = {978-1-4503-1321-6}
}

@article{ladner2015design,
  author = {Ladner, Richard E.},
  title = {Design for user empowerment},
  journal = {Interactions},
  volume = {22},
  number = {2},
  year = {2015},
  pages = {24--29},
  doi = {10.1145/2723869},
  publisher = {Association for Computing Machinery},
  address = {New York, NY, USA},
  issn = {1072-5520}
}

@inproceedings{bigham2017nkwydk,
  author    = {Bigham, Jeffrey P. and Lin, Irene and Savage, Saiph},
  title     = {The Effects of ``Not Knowing What You Don't Know'' on Web Accessibility for Blind Web Users},
  booktitle = {Proceedings of the 19th International ACM SIGACCESS Conference on Computers and Accessibility},
  series    = {ASSETS '17},
  year      = {2017},
  pages     = {101--109},
  publisher = {ACM},
  address   = {New York, NY, USA},
  doi       = {10.1145/3132525.3132533},
  url       = {https://doi.org/10.1145/3132525.3132533}
}

@article{BraunClarke2006,
  author  = {Braun, Virginia and Clarke, Victoria},
  title   = {Using thematic analysis in psychology},
  journal = {Qualitative Research in Psychology},
  year    = {2006},
  volume  = {3},
  number  = {2},
  pages   = {77--101},
  doi     = {10.1191/1478088706qp063oa},
  publisher = {Informa UK Limited}
}

@inproceedings{banovic2013uncovering,
  author    = {Banovic, Nikola and Franz, Rachel L. and Truong, Khai N. and Mankoff, Jennifer and Dey, Anind K.},
  title     = {Uncovering information needs for independent spatial learning for users who are visually impaired},
  booktitle = {Proceedings of the 15th International ACM SIGACCESS Conference on Computers and Accessibility (ASSETS '13)},
  year      = {2013},
  articleno = {24},
  numpages  = {8},
  pages     = {1--8},
  publisher = {Association for Computing Machinery},
  address   = {New York, NY, USA},
  isbn      = {9781450324052},
  doi       = {10.1145/2513383.2513445}
}

@inproceedings{poppinga2011touchover,
  author    = {Poppinga, Benjamin and Magnusson, Charlotte and Pielot, Martin and Rassmus-Gr{\"o}hn, Kirsten},
  title     = {TouchOver map: audio-tactile exploration of interactive maps},
  booktitle = {Proceedings of the 13th International Conference on Human Computer Interaction with Mobile Devices and Services (MobileHCI '11)},
  year      = {2011},
  pages     = {545--550},
  publisher = {ACM},
  address   = {New York, NY, USA},
  doi       = {10.1145/2037373.2037458}
}

@article{espinosa1998comparing,
  author  = {Espinosa, M. Angeles and Ungar, Simon and Ocha{\'i}ta, Esperanza and Blades, Mark and Spencer, Christopher},
  title   = {Comparing Methods for Introducing Blind and Visually Impaired People to Unfamiliar Urban Environments},
  journal = {Journal of Environmental Psychology},
  volume  = {18},
  number  = {3},
  pages   = {277--287},
  year    = {1998},
  doi     = {10.1006/jevp.1998.0097}
}

@article{lahav2008haptic,
  author  = {Orly Lahav and David Mioduser},
  title   = {Haptic-feedback support for cognitive mapping of unknown spaces by people who are blind},
  journal = {International Journal of Human-Computer Studies},
  volume  = {66},
  number  = {1},
  pages   = {23--35},
  year    = {2008},
  doi     = {10.1016/j.ijhcs.2007.08.001}
}

@inproceedings{abdolrahmani2017embracing,
  author    = {Abdolrahmani, Ali and Easley, William and Williams, Michele and Branham, Stacy and Hurst, Amy},
  title     = {Embracing Errors: Examining How Context of Use Impacts Blind Individuals' Acceptance of Navigation Aid Errors},
  booktitle = {Proceedings of the 2017 CHI Conference on Human Factors in Computing Systems},
  series    = {CHI '17},
  year      = {2017},
  pages     = {4158--4169},
  publisher = {Association for Computing Machinery},
  address   = {New York, NY, USA},
  isbn      = {978-1-4503-4655-9},
  doi       = {10.1145/3025453.3025528}
}

@inproceedings{rowell2005feeling,
  author    = {Rowell, Jonathan and Ungar, Simon},
  title     = {Feeling Our Way: Tactile Map User Requirements - A Survey},
  booktitle = {Proceedings of the 22nd International Cartographic Conference (ICC 2005)},
  address   = {A Coru\~{n}a, Spain},
  year      = {2005},
  publisher = {International Cartographic Association},
  url       = {https://icaci.org/files/documents/ICC_proceedings/ICC2005/htm/pdf/oral/TEMA22/Session\%201/JONATHAN\%20ROWELL.pdf}
}

@misc{applemapsvoiceover,
  title = {Use VoiceOver in apps on iPhone},
  author = {{Apple Inc.}},
  howpublished = {iPhone User Guide (iOS 26), \url{https://support.apple.com/guide/iphone/use-voiceover-in-apps-iphe4ee74be8/ios}},
  year = {2026},
  note = {Section ``Navigate in Maps''. Checked 30 August 2026}
}

@misc{voicevista,
  title = {VoiceVista: Blind Navigation},
  author = {Wu, Jianfeng},
  howpublished = {\url{https://drwjf.github.io/vvt/}},
  year = {2026},
  note = {Built on Microsoft's open-sourced Soundscape under the MIT License. Checked 30 August 2026}
}

@article{lederman1987hand,
  title = {Hand Movements: A Window into Haptic Object Recognition},
  author = {Lederman, Susan J. and Klatzky, Roberta L.},
  journal = {Cognitive Psychology},
  volume = {19},
  number = {3},
  pages = {342--368},
  year = {1987},
  doi = {10.1016/0010-0285(87)90008-9}
}

@article{scaife1996external,
  title = {External cognition: how do graphical representations work?},
  author = {Scaife, Mike and Rogers, Yvonne},
  journal = {International Journal of Human-Computer Studies},
  volume = {45},
  number = {2},
  pages = {185--213},
  year = {1996},
  doi = {10.1006/ijhc.1996.0048}
}

@inproceedings{hong2024errors,
  title = {Understanding How Blind Users Handle Object Recognition Errors: Strategies and Challenges},
  author = {Hong, Jonggi and Kacorri, Hernisa},
  booktitle = {Proceedings of the 26th International ACM SIGACCESS Conference on Computers and Accessibility (ASSETS '24)},
  pages = {1--15},
  year = {2024},
  publisher = {ACM},
  doi = {10.1145/3663548.3675635}
}

@article{kaplan2024audiohaptic,
  title   = {Fully Digital Audio Haptic Maps for Individuals with Blindness},
  author  = {Kaplan, Howard and Pyayt, Anna},
  journal = {Disabilities},
  volume  = {4},
  number  = {1},
  pages   = {64--77},
  year    = {2024},
  doi     = {10.3390/disabilities4010005}
}

@inproceedings{biggs2022evaluation,
  title     = {Evaluation of a Non-Visual Auditory Choropleth and Travel Map Viewer},
  author    = {Biggs, Brandon and Toth, Christopher and Stockman, Tony and Coughlan, James M. and Walker, Bruce N.},
  booktitle = {Proceedings of the International Conference on Auditory Display (ICAD 2022)},
  pages     = {82--90},
  year      = {2022},
  doi       = {10.21785/icad2022.027}
}

@article{giudice2020cognitive,
  title   = {Cognitive Mapping Without Vision: Comparing Wayfinding Performance After Learning From Digital Touchscreen-Based Multimodal Maps vs. Embossed Tactile Overlays},
  author  = {Giudice, Nicholas A. and Guenther, Benjamin A. and Jensen, Nicholas A. and Haase, Kaitlyn N.},
  journal = {Frontiers in Human Neuroscience},
  volume  = {14},
  pages   = {87},
  year    = {2020},
  doi     = {10.3389/fnhum.2020.00087}
}

@inproceedings{zhao2024tada,
  title     = {TADA: Making Node-link Diagrams Accessible to Blind and Low-Vision People},
  author    = {Zhao, Yichun and Nacenta, Miguel A. and Sukhai, Mahadeo A. and Somanath, Sowmya},
  booktitle = {Proceedings of the 2024 CHI Conference on Human Factors in Computing Systems},
  year      = {2024},
  doi       = {10.1145/3613904.3642222},
  numpages= {20},
}

@inproceedings{nagassa2023building,
  title     = {3D Building Plans: Supporting Navigation by People who are Blind or have Low Vision in Multi-Storey Buildings},
  author    = {Nagassa, Ruth G. and Butler, Matthew and Holloway, Leona and Goncu, Cagatay and Marriott, Kim},
  booktitle = {Proceedings of the 2023 CHI Conference on Human Factors in Computing Systems},
  year      = {2023},
  doi       = {10.1145/3544548.3581389},
  numpages= {19},
}

@article{belkin1995cases,
  title = {Cases, Scripts, and Information-Seeking Strategies: On the Design of Interactive Information Retrieval Systems},
  author = {Belkin, Nicholas J. and Cool, Colleen and Stein, Adelheit and Thiel, Ulrich},
  journal = {Expert Systems with Applications},
  volume = {9},
  number = {3},
  pages = {379--395},
  year = {1995},
  doi = {10.1016/0957-4174(95)00011-W}
}

@article{simonnet2019comparing,
  title   = {Comparing Interaction Techniques to Help Blind People Explore Maps on Small Tactile Devices},
  author  = {Simonnet, Mathieu and Brock, Anke M. and Serpa, Antonio and Oriola, Bernard and Jouffrais, Christophe},
  journal = {Multimodal Technologies and Interaction},
  volume  = {3},
  number  = {2},
  pages   = {27},
  year    = {2019},
  doi     = {10.3390/mti3020027},
  publisher = {MDPI}
}

@article{kitchin1997understanding,
  author  = {Kitchin, Robert M. and Blades, Mark and Golledge, Reginald G.},
  title   = {Understanding spatial concepts at the geographic scale without the use of vision},
  journal = {Progress in Human Geography},
  volume  = {21},
  number  = {2},
  pages   = {225--242},
  year    = {1997},
  doi     = {10.1191/030913297668904166}
}

@inproceedings{rowell2003feelingyour,
  author    = {Rowell, Jonathan and Ungar, Simon},
  title     = {Feeling Your Way: A Tactile Map User Survey},
  booktitle = {Proceedings of the 21st International Cartographic Conference (ICC): Cartographic Renaissance},
  address   = {Durban, South Africa},
  year      = {2003},
  pages     = {652--659},
  isbn      = {0-958-46093-0},
  publisher = {International Cartographic Association}
}

@inproceedings{kane2013touchplates,
  author    = {Kane, Shaun K. and Morris, Meredith Ringel and Wobbrock, Jacob O.},
  title     = {Touchplates: Low-Cost Tactile Overlays for Visually Impaired Touch Screen Users},
  booktitle = {Proceedings of the 15th International ACM SIGACCESS Conference on Computers and Accessibility (ASSETS '13)},
  year      = {2013},
  publisher = {ACM},
  address   = {New York, NY, USA},
  doi       = {10.1145/2513383.2513442},
  numpages= {8},
}

@article{cole2021tactile,
  author  = {Cole, Harrison},
  title   = {Tactile cartography in the digital age: A review and research agenda},
  journal = {Progress in Human Geography},
  volume  = {45},
  number  = {4},
  pages   = {834--854},
  year    = {2021},
  doi     = {10.1177/0309132521995877}
}

@article{reinders2024rtd,
  author  = {Reinders, Samuel and Butler, Matthew and Zukerman, Ingrid and Lee, Bongshin and Qu, Lizhen and Marriott, Kim},
  title   = {When Refreshable Tactile Displays Meet Conversational Agents: Investigating Accessible Data Presentation and Analysis with Touch and Speech},
  journal = {IEEE Transactions on Visualization and Computer Graphics},
  year    = {2025},
  doi     = {10.1109/TVCG.2024.3456358},
  note    = {IEEE VIS 2024, Honorable Mention Award},
  pages   = {864--874},
  volume  = {31},
  number  = {1},
}

@article{larkin1987diagram,
  title = {Why a Diagram is (Sometimes) Worth Ten Thousand Words},
  author = {Larkin, Jill H. and Simon, Herbert A.},
  journal = {Cognitive Science},
  volume = {11},
  number = {1},
  pages = {65--99},
  year = {1987},
  doi = {10.1111/j.1551-6708.1987.tb00863.x}
}

@article{hollan2000distributed,
  title = {Distributed Cognition: Toward a New Foundation for Human-Computer Interaction Research},
  author = {Hollan, James and Hutchins, Edwin and Kirsh, David},
  journal = {ACM Transactions on Computer-Human Interaction},
  volume = {7},
  number = {2},
  pages = {174--196},
  year = {2000}
}

@article{boeing2017osmnx,
  title = {OSMnx: New methods for acquiring, constructing, analyzing, and visualizing complex street networks},
  author = {Boeing, Geoff},
  journal = {Computers, Environment and Urban Systems},
  volume = {65},
  pages = {126--139},
  year = {2017},
  doi = {10.1016/j.compenvurbsys.2017.05.004}
}

@article{rieser1980role,
  author  = {Rieser, John J. and Lockman, Jeffrey J. and Pick, Herbert L.},
  title   = {The role of visual experience in knowledge of spatial layout},
  journal = {Perception \& Psychophysics},
  volume  = {28},
  number  = {3},
  pages   = {185--190},
  year    = {1980},
  doi     = {10.3758/BF03204374}
}

@incollection{siegel1975development,
  author    = {Siegel, Alexander W. and White, Sheldon H.},
  title     = {The Development of Spatial Representations of Large-Scale Environments},
  booktitle = {Advances in Child Development and Behavior},
  volume    = {10},
  pages     = {9--55},
  publisher = {Academic Press},
  year      = {1975}
}

@incollection{montello1998framework,
  author    = {Montello, Daniel R.},
  title     = {A New Framework for Understanding the Acquisition of Spatial Knowledge in Large-Scale Environments},
  booktitle = {Spatial and Temporal Reasoning in Geographic Information Systems},
  editor    = {Egenhofer, Max J. and Golledge, Reginald G.},
  publisher = {Oxford University Press},
  address   = {New York},
  pages     = {143--154},
  year      = {1998}
}

@inproceedings{wilson2007swan,
  author    = {Wilson, Jeff and Walker, Bruce N. and Lindsay, Jeffrey and Cambias, Craig and Dellaert, Frank},
  title     = {{SWAN}: System for Wearable Audio Navigation},
  booktitle = {Proceedings of the 11th IEEE International Symposium on Wearable Computers (ISWC '07)},
  year      = {2007},
  pages     = {91--98},
  publisher = {IEEE},
  doi       = {10.1109/ISWC.2007.4373786}
}

\onecolumn
\appendix

\section{Co-design record}
\label{app:codesign}

The co-designer (P01) wrote to us throughout the project, in English and in Chinese; Chinese passages are given in our translation. Personal details and quotations appear with their consent. The table lists the exchanges that bear on the design, in order, and what each one gave it. Dates are given to the month.

On the layout a route is chosen from, in the second pilot session:

\begin{quote}
\emph{Because to a blind person --- not only the route.\ldots{} And to me, before I know the route, I need to have a two-dimensional map. What I'm spending time to say here could be helpful to you when you design your product. This is how a blind person tries to understand the whole world. So I need to have a 2D, two-dimensional to the map. Like you guys, if you want to drive to Seattle --- back in the old days, we would have gone to AAA to get a map of the West Coast. Paper map. Highway 1 now to Highway 5\ldots{} then turn on to whatever that state highway. That would be your approach. So to me, the same thing. I need to have a map before I decide, determine the route.}
\end{quote}

{\small
\begin{longtable}{p{0.14\linewidth} p{0.49\linewidth} p{0.31\linewidth}}
\caption{Exchanges with the co-designer that bear on the design, from the email record and meeting notes, with contact details and third parties removed.}
\Description{A three-column table listing exchanges with the co-designer. Each row gives when the exchange happened, what was said or done, and what it gave the design.}\label{tab:codesign}\\
\toprule
When & What was said or done & What it gave the design \\
\midrule
\endfirsthead

\multicolumn{3}{l}{\textit{Table \thetable\ continued from previous page}}\\
\toprule
When & What was said or done & What it gave the design \\
\midrule
\endhead

\midrule
\multicolumn{3}{r}{\textit{Continued on next page}}\\
\endfoot

\bottomrule
\endlastfoot

2025-04 & First contact. They described themselves as a person who forms mental maps to reach destinations, and said recent hearing loss had made a familiar college campus hard to navigate. & The need: a spatial channel that does not depend on hearing (DG5). \\
\midrule
2025-07 & They inventoried how they orient: a shoreline followed with the cane, slope and surface under the feet, the sun, smells, sounds. On a loop between two places on campus, picnic tables turned them around; they reused this loop to test every assistant. & The recurring case that a point-to-point route without an area around it fails. \\
\midrule
2025-07 to 2025-08 & They tested a live camera assistant and wrote after each walk. It gave left and right, step counts and names; it also drifted, described what was no longer in view, confused a fountain with a sunken garden, and could not run alongside a map application. & Formative findings: the assistant answers about the moment and flips directions; nothing holds the answers together. \\
\midrule
2025-08 & On a campus walk we traced a plaza on their palm. They wrote that it overturned the mental map they had held of the place, and that the live camera assistant gives only ``the line between point and point, and hardly the area''; they called that blind spot fatal. They asked for an aerial plan of a bounded area, one they could read without first knowing their own position in it. & The map itself: an area under the hand, with the finger's position a place on it (DG3). \\
\midrule
2025-08 & Their rule for any description or tactile object: the rough outline first, then detail; a good-enough rendering over a faithful one, since too much information is hard to process. & Staged disclosure (DG2) and the choice of simplified footprints. \\
\midrule
2025-08 & They described their checking practice: test the machine on things they know, such as a fence height or a step count, and hold the machine's account and their memory both open when they disagree. They noted that people blind from birth cannot check a description against visual memory. & The check-by-hand theme, and its limit. \\
\midrule
2026-02 to 2026-04 & They asked for 3D printed tactile maps of three places they already knew so that they could read them against memory. The first print carried no names; the next carried braille labels. & Reading a map against memory as a task; name on entry as the spoken equivalent of a label. \\
\midrule
2026-05 & Pilot 1 on the working prototype. They could not build a stable picture from a relative pointer and asked for the map to be almost the size of the trackpad; taps and resting fingers registered as clicks. & Absolute mapping, click-free pad, three physical keys, onboarding rewritten into short lines with a practice square. \\
\midrule
2026-06 & Pilot 2 with the final configuration. They distinguished a generative assistant, a consultant, from an agent, a secretary, and for a blind walker an electronic guide dog that warns of a pillar ahead. Their own protocol with any assistant was overview first, what is east and west and how far, then asking on approach. & Final pre-study configuration; their framing of what the agent should be. \\
\end{longtable}
}

\end{document}